\documentclass[a4paper]{article}

\usepackage{makecell}
\usepackage[utf8]{inputenc}
\usepackage[top=10mm]{geometry}
\usepackage{booktabs}
\usepackage[flushleft]{threeparttable}
\usepackage{authblk}
\usepackage{bm}
\usepackage{amsmath}
\usepackage{amssymb}
\usepackage{graphicx}
\usepackage[square,sort,numbers,compress]{natbib}
\usepackage{slashed}
\usepackage{float}
\usepackage{longtable,booktabs}
\usepackage[colorlinks,
linkcolor=blue,
anchorcolor=blue,
urlcolor=blue,
citecolor=blue
]{hyperref}
 \allowdisplaybreaks[4]

\title{\large \textbf{Pion-nucleon scattering with decuplet contribution in heavy baryon SU(3) chiral perturbation theory}}

\author[1]{\small Jing Ou-Yang }
\author[1,2]{\small Bo-Lin Huang \thanks{blhuang@imu.edu.cn }}

\affil[1]{\textit{\small School of Physical Science and Technology, Inner Mongolia University, Hohhot 010021, China}}
\affil[2]{\textit{\small Center for Quantum Physics and Technologies, School of Physical Science and Technology, Inner Mongolia University, Hohhot 010021, China}}

\date{\small \today}
\begin{document}
\maketitle

%abstract
\begin{abstract}
We calculate the complete $T$ matrices with decuplet contributions for pion-nucleon scattering to order $\mathcal{O}(\epsilon^3)$ in heavy baryon SU(3) chiral perturbation theory. The baryon mass in the chiral limit $M_0$ and the low-energy constants are determined by fitting to phase shifts of $\pi N$, the experimental octet-baryon masses, and the value of $\sigma_{\pi N}$ simultaneously. By using these constants, we obtain the $KN$ $\sigma$ terms, $\sigma_{KN}^{(1)}=(375.07\pm33.02)$ MeV and $\sigma_{KN}^{(2)}=(275.32\pm32.24)$ MeV, with the errors being only statistical. An excellent description of the phase shifts is obtained for all partial waves. We also present results for scattering lengths and scattering volumes. In addition, the convergence of the approach is also discussed.
\end{abstract}

\section{Introduction}

The scattering process between pions and nucleons is crucial for a deep understanding of chiral dynamics in quantum chromodynamics (QCD). %This process has been investigated up to the third and fourth orders in SU(3) framework, thereby providing precise predictions for the threshold parameters\cite{huan20201,huan20202}. 
%This methodology is based on an effective field theory, where the active degrees of freedom consist of asymptotically observable pion and nucleon fields. 
Chiral perturbation theory (ChPT) \cite{wein1979,gass1984,leut1994,sche2012}, as an effective field theory of QCD at energies below the chiral symmetry breaking scale $\Lambda_\chi$ of approximately 1 GeV, provides a suitable framework for calculating the model-independent pion-nucleon ($\pi N$) scattering amplitude. 

In recent years, significant advances have been made in the study of ChPT. %in the contexts of flavor SU(2) \cite{gass1984} and flavor SU(3) \cite{gass1985}. 
These advancements have extended into the domain of ChPT that includes baryons \cite{meis1993}. Although the theory has achieved some success in areas that include baryons, the non-vanishing mass of baryon in the chiral limit poses challenges. % The relativistic framework within ChPT does not naturally provide a power counting rule due to the nucleon mass acting as a new energy scale in the chiral limit. 
The baryon mass introduces a new energy scale in the chiral limit, so the relativistic frame of ChPT does not naturally provide a power counting rule. In 1988, Gasser first applied baryon ChPT to elastic $\pi N$ scattering \cite{gass1988}, but the power counting rules for the baryonic sector were invalidated due to the nucleon mass in the chiral limit. To address this issue, various methods have been proposed, including relativistic approaches such as infrared regularization \cite{bech1999} and the extended on-mass-shell scheme \cite{gege1999,fuch2003}, as well as non-relativistic methods, specifically heavy baryon chiral perturbation theory (HBChPT) \cite{jenk1991,bern1992}. Relativistic approaches have made substantial progress in many aspects \cite{Schindler:2006it,geng2008,MartinCamalich:2010fp,Alarcon:2011zs,chen2013,Yao:2016vbz,lu2019}. However, HBChPT has played a crucial role in the study of low-energy meson-baryon interactions. By employing non-relativistic methods, we can calculate not only non-relativistic scattering amplitudes but also account for relativistic contributions by including corrections of $1/M_0$, where $M_0$ represents the baryon mass in the chiral limit.

With increasing energy levels, the applicability of chiral expansion encounters limitations due to the manifestation of nucleon resonances, particularly the $\Delta$(1232) resonance, distinguished by its spin and isospin of 3/2, which is of paramount significance. The relevance of this resonance to hadronic and nuclear physics has been well validated. %In line with the foundational insights of Jenkins and Manohar \cite{jenk1991}, there exists a coherent and methodical framework that integrates this critical degree of freedom within baryon ChPT, an endeavor elucidated by Hemmert, Holstein, and Kambor \cite{hemm1998}. 
In accordance with the foundational insights of Jenkins and Manohar \cite{jenk1991}, there is a coherent and systematic framework that incorporates this crucial degree of freedom within baryon ChPT, as further elucidated by Hemmert, Holstein, and Kambor \cite{hemm1998}. %Considering the Nucleon–Delta mass splitting as an additional small parameter facilitates the derivation of what is termed the small scale expansion (SSE), distinct from the chiral expansion due to the persistence of the $N\Delta$ splitting at the chiral limit. 
Considering the Nucleon–Delta mass splitting as an additional small parameter enables the derivation of the small scale expansion (SSE), which is distinct from the chiral expansion due to the persistence of the $N\Delta$ splitting even in the chiral limit.
The SSE refers to the expansion in terms of small quantities such as $p/\Lambda$, $m/\Lambda$, or $\delta/\Lambda$, where $p$ and $m$ denote the momentum and mass of the meson, respectively, and $\delta$ represents the mass difference between the baryon decuplet and octet. %Prior research has substantiated that the delta resonance saturates most of the low-energy constants in the effective chiral pion–nucleon Lagrangian \cite{bern1997}.
It has been demonstrated that the $\Delta$ resonance accounts for most of the low-energy constants in the effective chiral pion–nucleon Lagrangian \cite{bern1997}. Consequently, the resummation of terms within the framework of the SSE yields improved convergence relative to the chiral expansion. %Furthermore, the inclusion of the delta as an explicit degree of freedom notably extends the radius of convergence of the series. This evidently enhances the intricacy of the methodology, as the relevant effective Lagrangian encompasses additional structures that are congruent with all underlying symmetries.

Within the SSE framework, pion-nucleon scattering involving $\Delta$ resonance has been thoroughly analyzed using both relativistic \cite{Yao:2016vbz} and non-relativistic \cite{fett2001} SU(2) ChPT. In both cases, an accurate description was achieved for the phase shifts, particularly for the $P_{33}$ partial wave. Furthermore, the convergence had also been significantly improved. The crucial $\sigma_{\pi N}$ value could also be obtained as a byproduct. However, for processes involving kaons or hypersons, one has to use three-flavor chiral dynamics. The analysis of meson-baryon scattering with SU(3) ChPT has proven to be effective \cite{kais2001,liu20071,huan2015,huan2017}. In particular, our previous work on pion-nucleon scattering has also achieved good results \cite{huan20201,huan20202}. A more accurate result can be achieved by considering baryon resonances, similar to the SU(2) case. The meson-baryon scattering lengths were obtained by including decuplet contributions, as detailed in ref.~\cite{liu20072}. In this paper, we will calculate the complete $T$ matrices with decuplet contributions for pion-nucleon scattering to third order in SU(3) HBChPT. The baryon mass in the chiral limit $M_0$ and the low-energy constants will be determined by fitting to phase shifts of pion-nucleon, the experimental octet-baryon masses, and the value of $\sigma_{\pi N}$ simultaneously. The results are accurate because precise data was used for the fitting. Then, the $KN$ $\sigma$ terms will be predicted with these constants. Similar to the $\sigma_{\pi N}$ value, they might be very helpful for the direct detection of dark matter \cite{hofe2015,cush2013}. Therefore, our calculation of pion-nucleon scattering with decuplet contributions in SU(3) HBChPT is interesting.

The structure of the paper is organized as follows. In Sec.~\ref{lagrangian}, we introduce the Lagrangians. In Sec.~\ref{tmatrices}, we present the explicit results of $T$-matrices with decuplet contributions. In Sec.~\ref{phase}, we outline how to calculate partial-wave phase shifts, scattering lengths and scattering volumes. In Sec.~\ref{baryonmass}, we explain how to calculate the baryon masses and the $\sigma$-terms with decuplet contributions. Section~\ref{results} contains the results and discussion and also includes a brief summary. The Appendix contains the one-loop amplitudes with decuplet contribution.

\section{Chiral Lagrangian}
\label{lagrangian}
In order to calculate the pion-nucleon scattering amplitudes with decuplet baryons in heavy baryon SU(3) chiral perturbation theory, the corresponding effective Lagrangian has the form
\begin{align}
\label{eq1}
\mathcal{L}_{\text{eff}}=\mathcal{L}_{\phi\phi}+\mathcal{L}_{\phi B}+\mathcal{L}_{\phi B T}.
\end{align}
The traceless Hermitian $3\times 3$ matrices $\phi$ and $B$ include the pseudoscalar Goldstone boson fields ($\pi$, $K$, $\bar{K}$, $\eta$) and
the octet-baryon fields ($N$, $\Lambda$, $\Sigma$, $\Xi$), respectively. The $T$ represents the decuplet-baryon fields($\Delta,\Sigma^*,\Xi^*,\Omega$). The lowest-order SU(3) chiral Lagrangians of the three parts take the form \cite{liu20072}
\begin{align}
\label{eq2}
\mathcal{L}^{(2)}_{\phi\phi}=f^2\text{tr}(u_\mu u^\mu +\frac{\chi_{+}}{4}),
\end{align}
\begin{align}
\label{eq3}
 \mathcal{L}_{\phi B}^{(1)}=\text{tr}(i\overline{B}[v\cdot D,B])+2D\,\text{tr}(\overline{B}S_{\mu}\{u^{\mu},B\})+2F\,\text{tr}(\overline{B}S_{\mu}[u^{\mu},B]),
\end{align}
\label{eq4}
\begin{align}
\mathcal{L}_{
\phi B T}^{(1)}=-\bar{T}^{\mu}(i v\cdot D-\delta)T_\mu+\mathcal{C}(\bar{T}^\mu u_\mu B+\bar{B}u_\mu T^{\mu})+2\mathcal{H}\bar{T}^{\mu}S\cdot u T_\mu,
\end{align}
where $f$ is the pseudoscalar decay constant in the chiral limit. The axial vector quantity $u^\mu=i\{\xi^{\dagger},\partial^\mu\xi\}/2$ contains odd number meson fields. The quantity $\chi_{+}=\xi^{\dagger}\chi\xi^{\dagger}+\xi\chi\xi$ with $\chi=\text{diag}(m_\pi^2,m_\pi^2,2m_K^2-m_\pi^2)$ introduces explicit chiral symmetry breaking terms. The superscripts in these Lagrangians
denote the order of the small scale expansion. We choose the SU(3) matrix
\begin{align}
\label{eq5}
U=\xi^2=\text{exp}(i\phi/f),
\end{align}
which collects the pseudoscalar Goldstone boson fields. Note that, the so-called sigma parametrization was chosen in SU(2) HB$\chi$PT \cite{mojz1998,fett1998}. The $D_{\mu}$ denotes the chiral covariant derivative
\begin{align}
\label{eq6}
[D_{\mu},B]=\partial_{\mu}B+[\Gamma_{\mu},B],
\end{align}
and $S_{\mu}$ is the covariant spin operator
\begin{align}
\label{eq7}
S_\mu=\frac{i}{2}\gamma_5 \sigma_{\mu\nu}v^\nu,\quad S\cdot v=0,
\end{align}
\begin{align}
\label{eq8}
\{S_\mu,S_\nu\}=\frac{1}{2}(v_\mu v_\nu-g_{\mu\nu}),\quad [S_\mu,S_\nu]=i\epsilon_{\mu\nu\sigma\rho}v^\sigma S^\rho,
\end{align}
where $\epsilon_{\mu\nu\sigma\rho}$ is the completely antisymmetric tensor in four indices, $\epsilon_{0123}=1.$ The chiral connection $\Gamma^\mu=[\xi^{\dagger},\partial^\mu\xi]/2$ contains even number meson fields. The physical values of the axial vector coupling constants $D$ and $F$ can be determined in fits to semileptonic hyperon decays \cite{bora1999}. The physical values of the axial-current couplings $\mathcal{C}$ and $\mathcal{H}$ can be obtained from the strong and electromagnetic decays of the decuplet baryons \cite{butl1992}. There are relations: $D+F=g_A$, $\mathcal{C}=\sqrt{2}g_{\pi N \Delta}$ and $\mathcal{H}=g_1$ in comparison with the SU(2) case of ref.~\cite{fett2001}. The $\delta$ denotes the mass difference between decuplet and octet baryon in the chiral limit. The chiral covariant derivative with decuplet baryon field reads
\label{eq9}
\begin{align}
iD_\mu{T_{abc}^\nu}=i\partial_\mu{T_{abc}^\nu}+(\Gamma_\mu)_a^d{T_{dbc}^\nu}+(\Gamma_\mu)_b^d{T_{adc}^\nu}+(\Gamma_\mu)_c^d{T_{abd}^\nu},
\end{align}
with 
\label{eq10}
\begin{align}
T_{abc}=\left(\left(
\begin{array}{ccc}
\Delta^{++} & \frac{\Delta^{+}}{\sqrt{3}} & \frac{\Sigma^{*+}}{\sqrt{3}} \\\\
\frac{\Delta^{+}}{\sqrt{3}} & \frac{\Delta^{0}}{\sqrt{3}} & \frac{\Sigma^{*0}}{\sqrt{6}} \\\\
\frac{\Sigma^{*+}}{\sqrt{3}} & \frac{\Sigma^{*0}}{\sqrt{6}} & \frac{\Xi^{*0}}{\sqrt{3}} \end{array}\right),\left(
\begin{array}{ccc}
\frac{\Delta^{+}}{\sqrt{3}} & \frac{\Delta^{0}}{\sqrt{3}} & \frac{\Sigma^{*0}}{\sqrt{6}} \\\\
\frac{\Delta^{0}}{\sqrt{3}}  & \Delta^{-}  & \frac{\Sigma^{*-}}{\sqrt{3}} \\\\
\frac{\Sigma^{*0}}{\sqrt{6}} & \frac{\Sigma^{*-}}{\sqrt{3}} & \frac{\Xi^{*-}}{\sqrt{3}} \end{array}\right),\left(
\begin{array}{ccc}
\frac{\Sigma^{*+}}{\sqrt{3}}  & \frac{\Sigma^{*0}}{\sqrt{6}}  & \frac{\Xi^{*0}}{\sqrt{3}} \\\\
\frac{\Sigma^{*0}}{\sqrt{6}}  & \frac{\Sigma^{*-}}{\sqrt{3}}  & \frac{\Xi^{*-}}{\sqrt{3}} \\\\
\frac{\Xi^{*0}}{\sqrt{3}} & \frac{\Xi^{*-}}{\sqrt{3}} & \Omega^{-} \end{array}\right)\right).
\end{align}
In this representation, one can assign any particular permutation of indices $a$, $b$, $c$ to denote the row, column, and sub-matrix. This is because the decuplet remains completely symmetric even after rearranging the flavor indices.

In our calculation of decuplet contributions to pion-nucleon scattering, we truncate at $\mathcal{O}(\epsilon^3)$. The explicit form of the $\mathcal{L}_{\phi B}^{(2)}$ and $\mathcal{L}_{\phi B}^{(3)}$ can be found in ref.~\cite{huan20201}, where the low-energy constants $b_i(i=D,F,0,1,...,11)$ and $h_i(i=1,...,13)$ are involved. The $h_i$ can absorb the divergences from loop calculations. The chiral Lagrangians $\mathcal{L}_{\phi B T}^{(2)}$ and $\mathcal{L}_{\phi B T}^{(3)}$ can be constructed by following ref.~\cite{fett2001}. However, the amplitudes from the tree diagrams involved the two Lagrangians are consistent with the ones of the SU(2) case. For this reason the notation of low-energy constants $b_3+b_8$, $f_1+f_2-\frac{b_8}{2m}$, $2f_4-f_5-\frac{b_8}{4m}$ (replaced by $C^T$, $H^T_1$, $H^T_2$ in this paper) for the second- and third-order counterterms in SU(2) case introduced in ref.~\cite{fett2001} will be used.

\section{$T$-matrices with decuplet contributions}
\label{tmatrices}
We are considering in this work only elastic pion-nucleon scattering processes $\pi(q)+N(p) \rightarrow \pi(q')+N(p')$ in center-of-mass system (CMS). In the total isospin  $I=(1/2,3/2)$ of the pion-nucleon system, the corresponding $T$-matrix takes the following form:
\begin{align}
\label{eq11}
 T_{\pi N}^{(I)}=&V_{\pi N}^{(I)}(w,t)+i\bm{\sigma}\cdot(\bm{q}'\times\bm{q})W_{\pi N}^{(I)}(w,t)
\end{align}
with $w=v\cdot q=v\cdot q'$ the pion CMS energy, $t=(q'-q)^2$ the invariant momentum transfer squared and
\begin{align}
\label{eq12}
\bm{q}'^{2}=\bm{q}^2=\frac{(M_{N}^{2}-{m_{\pi}}^{2})^2}{4E_{cm}^2}-\frac{M_{N}^{2}+{m_{\pi}}^{2}}{2}+\frac{E_{cm}^2}{4}=\frac{M_{N}^{2}\bm{p}_{\text{lab}}^{2}}{m_{\pi}^{2}+M_{N}^{2}+2M_{N}\sqrt{m_{\pi}^{2}+\bm{p}_{\text{lab}}^{2}}},
\end{align}
with $E_{cm}$ the total energy in CMS and $\bm{p}_{\text{lab}}$ the momentum of the incident meson in the laboratory system. Furthermore, $V_{\pi N}^{(I)}(w,t)$ refers to the non-spin-flip pion-nucleon amplitude and $W_{\pi N}^{(I)}(w,t)$ refers to the spin-flip pion-nucleon amplitude.

\begin{figure}[t]
\centering
\includegraphics[height=5cm,width=12cm]{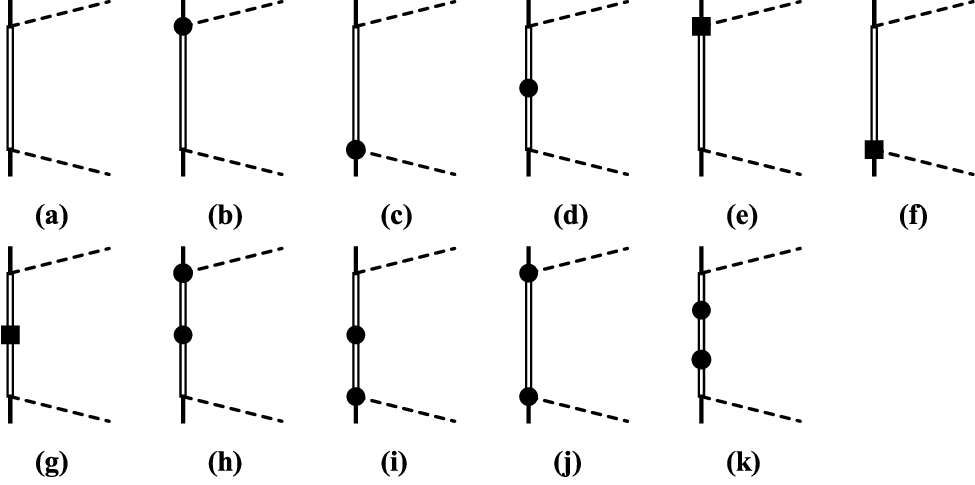}% Here is how to import EPS art
\caption{\label{fig:treedecupletfeynman}Tree diagrams with decuplet baryons. Dashed lines represent Goldstone bosons, solid lines represent octet baryons, and double-solid lines represent decuplet baryons. The heavy dots refer to vertices from the second order effective Lagrangian and the filled squares refer to insertions from the third order effective Lagrangian. Diagrams with crossed meson lines are not shown.}
\end{figure}

The $T$-matrices without decuplet contributions can be found in ref.~\cite{huan20201}. In the following, we calculate the $T$-matrices with decuplet contributions. The four-vector velocity is chosen as $v^{\mu}=(1,0,0,0)$ throughout this paper. The leading-order $\mathcal{O}(\epsilon)$ amplitudes with decuplet baryons resulting from Fig.~\ref{fig:treedecupletfeynman} read
\begin{align}
\label{eq13}
V_{\pi N}^{(3/2)}=\frac{\mathcal{C}^2(2\delta+w)}{9f^2(\delta^2-w^2)}(2w^2-2m_\pi^2+t),
\end{align}
\begin{align}
\label{eq14}
W_{\pi N}^{(3/2)}=-\frac{\mathcal{C}^2(\delta+2w)}{9f^2(\delta^2-w^2)},
\end{align}
\begin{align}
\label{eq15}
V_{\pi N}^{(1/2)}=\frac{2\mathcal{C}^2}{9f^2(\delta+w)}(2w^2-2m_\pi^2+t),
\end{align}
\begin{align}
\label{eq16}
W_{\pi N}^{(1/2)}=\frac{2\mathcal{C}^2}{9f^2(\delta+w)},
\end{align}
where $w=(m_\pi^2+\bm{q}^2)^{1/2}$ and $t=2\bm{q}^2(z-1)$ in the center-of-mass system with $z=\text{cos}(\theta)$ the cosine of the angle $\theta$ between $\bm{q}$ and $\bm{q}'$.  

\begin{figure}[t]
\centering
\includegraphics[height=16cm,width=14cm]{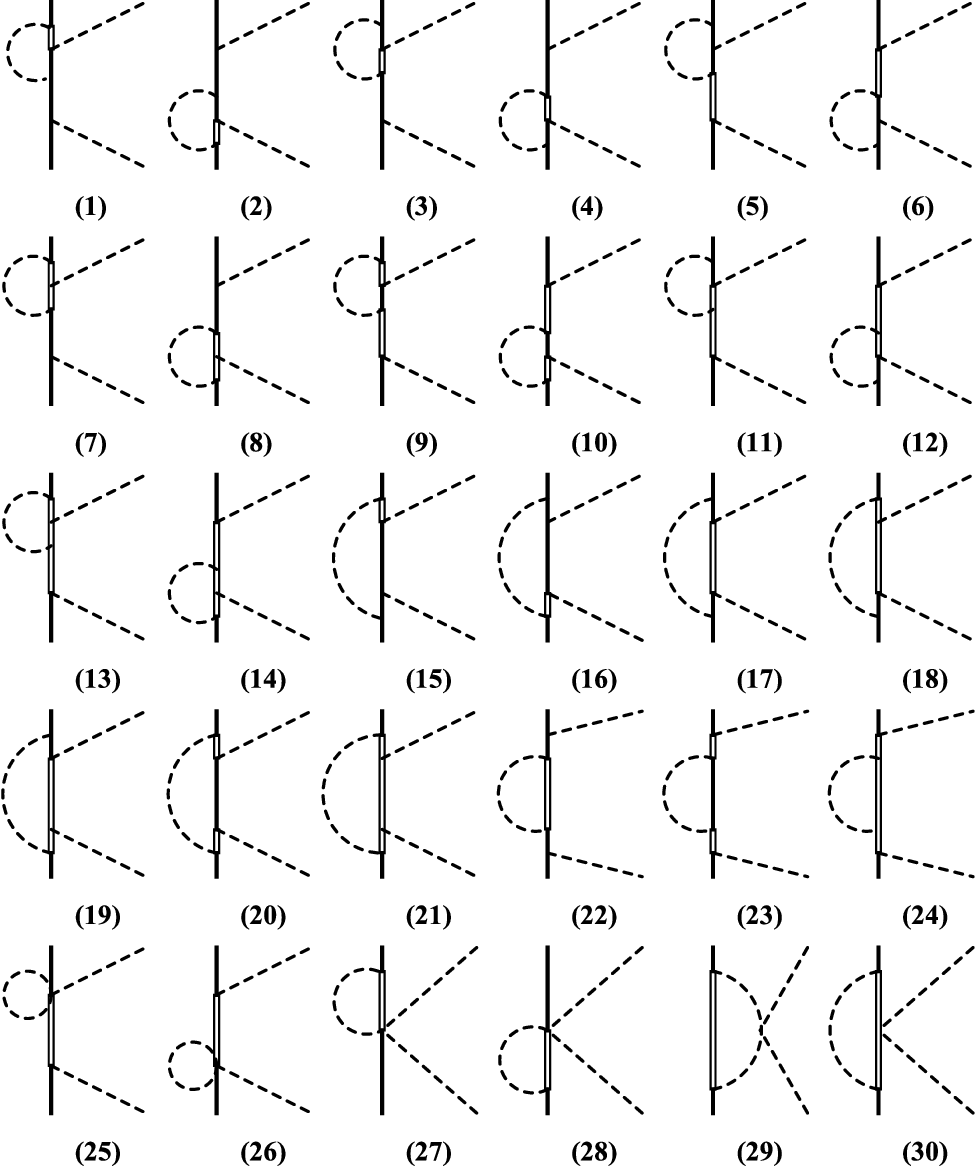}% Here is how to import EPS art
\caption{\label{fig:pinloopfeynman}Nonvanishing one-loop diagrams with decuplet baryons. The crossed-diagrams are not shown.  }
\end{figure}

At the second-order $\mathcal{O}(\epsilon^2)$ and the third-order $\mathcal{O}(\epsilon^3)$, the amplitudes without decuplet baryons are the same as the results from ref.~\cite{huan20201}, where the linear combinations of the low-energy constants $C_{1,2,3}$ and $H_{1,2,3,4}$ are also used in this paper. Meanwhile, the amplitudes with decuplet baryons can be obtained from the tree and one-loop diagrams in Figs.~\ref{fig:treedecupletfeynman}-\ref{fig:pinloopfeynman}. The amplitudes from the tree diagrams are the same as the results of the SU(2) case, where the low-energy constants $C^T$, $H^T_1$, $H^T_2$ are used in this paper. The explicit tree amplitudes can be found in ref.~\cite{fett2001}. The one-loop diagrams involve vertices from the Lagrangians $\mathcal{L}_{\phi \phi}^{(2)}$, $\mathcal{L}_{\phi B}^{(1)}$ and $\mathcal{L}_{\phi B T}^{(1)}$. We use dimensional regularization method to evaluate divergent loop integrals \cite{hoof1979,bern1995,bouz2000,bouz2002,meng2019,wang2019}. The finite parts from the one-loop calculation can be found in the Appendix~\ref{oneloopamplitudes}.

\section{Partial-wave phase shifts and scattering lengths}
\label{phase}
The partial-wave amplitudes $f_{j}^{(I)}(\bm{q}^2)$, where $j=l\pm 1/2$ refers to the total angular momentum and $l$ to orbital angular momentum, are obtained from the non-spin-flip and spin-flip amplitudes by a projection:
\begin{align}
\label{eq17}
f_{l\pm 1/2}^{(I)}(\bm{q}^2)=\frac{M_{N}}{8\pi\sqrt{s}}\int_{-1}^{+1}dz\Big\{V_{\pi N}^{(I)}P_{l}(z)+\bm{q}^{2}W_{\pi N}^{(I)}[P_{l\pm 1}(z)-zP_{l}(z)]\Big\},
\end{align}
where $P_{l}(z)$ denotes the conventional Legendre polynomial. For the energy range considered in this paper, the phase shifts $\delta_{l\pm 1/2}^{(I)}$ are calculated by (also see Refs.~\cite{gass1991,fett1998})
\begin{align}
\label{eq18}
\delta_{l\pm 1/2}^{(I)}=\text{arctan}[|\bm{q}|\,\text{Re}\,f_{l\pm 1/2}^{(I)}(\bm{q}^2)].
\end{align}

The scattering lengths for s waves and the scattering volumes for p waves are obtained by approaching the threshold \cite{eric1988}
\begin{align}
\label{eq19}
a_{l\pm 1/2}^{(I)}=\lim\limits_{|\bm{q}| \rightarrow 0}\bm{q}^{-2l}f_{l\pm 1/2}^{(I)}(\bm{q}^2).
\end{align}

\section{Baryon masses and $\sigma$-terms with decuplet contributions}
\label{baryonmass}
The baryon masses and $\sigma$-terms have been investigated up to the fourth order in the HB$\chi$PT \cite{bern19931,bora1997} and the covariant baryon chiral perturbation theory \cite{ren2012}. However, for consistency in our calculation, we take the expressions of baryon masses and $\sigma$-terms with decuplet baryons from Ref.~\cite{bern19931} in which a complete calculation up to order $\mathcal{O}(\epsilon^3)$ was done by using HB$\chi$PT. At this order, the octet-baryon $M_B (B=N, \Lambda, \Sigma, \Xi)$ masses take the form
\begin{align}
\label{eq20}
M_B=M_0-\frac{1}{24\pi f^2}(\alpha^\pi_B m_\pi^3+\alpha_B^K m_K^3+\alpha_B^\eta m_\eta^3)+\gamma_B^D b_D+\gamma_B^F b_F-2b_0(m_\pi^2+2m_K^2)+ \Delta M_B,
\end{align}
 The numerical factors $\alpha_B^\pi$, $\alpha_B^K$, $\alpha_B^\eta$, $\gamma_B^D$ and $\gamma_B^F$ can be found in Eq.~(2.6) of ref.~\cite{bern19931}. The decuplet contributions to the octet masses can be given by
\begin{align}
\label{eq21}
\Delta M_B=\frac{\mathcal{C}^2}{24\pi^2 f^2}\Big[\beta^\pi_B H(m_\pi)+\beta^K_B H(m_K)+\beta^\eta_B H(m_\eta)\Big],
\end{align}
with coefficients
\begin{align}
\label{eq22}
{\beta}^\pi_N=&4,\quad \beta^K_N=1,\quad \beta^\eta_N=0;\quad 
\beta^{\pi}_\Sigma=\frac{2}{3},\quad \beta^K_{\Sigma}=\frac{10}{3},\quad \beta^{\eta}_{\Sigma}=1;\nonumber\\
\beta^{\pi}_{\Lambda}=&3,\quad \beta^K_{\Lambda}=2,\quad \beta^{\eta}_{\Lambda}=0;\quad 
\beta^{\pi}_{\Xi}=1,\quad \beta^K_{\Xi}=3,\quad \beta^\eta_{\Xi}=1,
\end{align}
and
\label{eq23}
\begin{equation}
H(m_\phi)=\delta^3\text{ln}\frac{2\delta}{m_\phi}+ m_\phi^2\delta\Big(\frac{3}{2}\text{ln}\frac{m_\phi}{\lambda}-1\Big)
-\left\{
\begin{array}{ll}
(\delta^2-m_\phi^2)^{3/2}\text{ln}\Bigg[\frac{\delta}{m_\phi}+\sqrt{\frac{\delta^2}{m_\phi^2}-1}\Bigg] &  m_\phi < \delta \\
\\
(m_\phi^2-\delta^2)^{3/2}\arccos{\frac{\delta}{m_\phi}}  &  m_\phi > \delta \\
\end{array} \right..
\end{equation}

The sigma terms are the scalar form factors of baryons which measure the strength of the various matrix elements $m_q\bar{q}q$ in the baryons. The nucleon sigma term with decuplet contribution at zero momentum transfer is given as \cite{bern19931}
\begin{align}
\label{eq24}
\sigma_{\pi N}=\frac{m_\pi^2}{64\pi f^2}\Big[-4\alpha^\pi_N m_\pi-2\alpha^K_N m_K-\frac{4}{3}\alpha^\eta_N m_\eta\Big]-2m_\pi^2(b_D+b_F+2b_0)+\Delta\sigma_{\pi N},
\end{align}
\begin{align}
\label{eq25}
\sigma_{K N}^{(j)}=&\frac{m_K^2}{64\pi f^2}\Bigg[-2\alpha^\pi_N m_\pi-3\xi^{(j)}_K m_K-\frac{10}{3}\alpha^\eta_N m_\eta-2\xi^{(j)}_{\pi \eta}\alpha^{\pi \eta}_N\frac{m_\pi^2+m_\pi m_\eta+m_\eta^2}{m_\pi+m_\eta}\Bigg]\nonumber\\&+4m_K^2(\xi^{(j)}_D b_D+\xi^{(j)}_F b_F-b_0)+\Delta\sigma_{K N}^{(j)},
\end{align}
\begin{align}
\label{eq25}
\sigma_{s N}=\Bigg(\frac{1}{2}-\frac{m_\pi^2}{4m_K^2}\Bigg)[3\sigma_{KN}^{(1)}+\sigma_{KN}^{(2)}]+\Bigg(\frac{1}{2}-\frac{m_K^2}{m_\pi^2}\Bigg)\sigma_{\pi N},
\end{align}
for $j = 1, 2$ with coefficients
\begin{align}
\label{26}
&\xi^{(1)}_K=\frac{7}{3}D^2-2DF+5F^2,\quad \xi^{(2)}_K=3(D-F)^2,\quad \xi^{(1)}_{\pi \eta}=1,\quad \xi^{(2)}_{\pi \eta}=-3,\quad \xi^{(1)}_D=-1,\nonumber\\&\xi^{(2)}_D=0,\quad \xi^{(1)}_F=0,\quad \xi^{(2)}_F=1;\quad \alpha^{\pi \eta}_N=\frac{1}{3}(D+F)(3F-D).
\end{align}
The decuplet contributes to the $\sigma$-terms are given by
\begin{align}
\label{eq27}
\Delta \sigma_{\pi N}=&\frac{m_\pi^2 \mathcal{C}^2}{64\pi^2 f^2}\Big[8\widetilde{H}(m_\pi)+\widetilde{H}(m_K)\Big],\nonumber\\
\Delta \sigma_{k N}^{(1)}=&\frac{m_K^2 \mathcal{C}^2}{64\pi^2 f^2}\Bigg[4\widetilde{H}(m_\pi)+\frac{4}{3}\widetilde{H}(m_K)\Bigg],\nonumber\\
\Delta \sigma_{k N}^{(2)}=&\frac{m_K^2 \mathcal{C}^2}{64\pi^2 f^2}\Big[4\widetilde{H}(m_\pi)+2\widetilde{H}(m_K)\Big],
\end{align}
with
\begin{equation}
\label{eq28}
\widetilde{H}(m_\phi)=\delta\Big(2\text{ln}{\frac{m_\phi}{\lambda}}-1\Big)+\left\{
\begin{array}{ll}
2\sqrt{\delta^2-m_\phi^2}\text{ln}\Bigg[\frac{\delta}{m_\phi}+\sqrt{\frac{\delta^2}{m_\phi^2}-1}\Bigg] &  m_\phi < \delta \\
\\
-2\sqrt{m_\phi^2-\delta^2}\arccos{\frac{\delta}{m_\phi}}  &  m_\phi > \delta \\
\end{array} \right..
\end{equation}

\section{Results and discussion}
\label{results}
Before making predictions, we have to determine the pertinent constants. Throughout this paper, we use $m_\pi=139.57$ MeV, $m_K=493.68$ MeV and $m_\eta=547.86 $MeV \cite{pdg2022}. For the pseudoscalar decay constants, we can either use $f_\pi=92.07$ MeV or $f_K=110.03$ MeV. To evaluate the sensitivity of our results to this higher-order effect, we can vary $f$ between $f_\pi$ and $f_K$. For simplicity, we adopt the average value $f=(f_\pi+f_K)/2=101.05$ MeV, as done in ref.~\cite{bern19931}.  Furthermore, we use $M_N =938.92\pm1.29$ MeV, $M_\Sigma=1191.01\pm4.86$ MeV, $M_\Xi=1318.26\pm6.30$ MeV and $M_\Lambda=1115.68\pm5.58$ MeV \cite{pdg2022}. Following ref.~\cite{liu20071}, we take the central value of $M_N$, $M_\Sigma$, and $M_\Xi$ to be the average of the isospin multiplet. Their error is simply the mass splitting of the isospin multiplet. The error of $M_\Lambda$ is added by approximately $0.5\%$ of the baryon mass because of the typical electromagnetic correction. We take $\delta=294$ MeV from the mass difference between the $N$ and the $\Delta$. We also set the scale $\lambda=4\pi f_\pi=1.16\,\text{GeV}$ as the chiral symmetry breaking scale. Recently, the $g_A$ was determined to be around 1.27 from the measurement in the decay of free neutrons \cite{mark2019} and the calculation in lattice quantum chromodynamics \cite{chan2018}. Thus, the $D=0.80$ and $F=0.47$ are taken as their physical values. We take $\mathcal{C}=-1.2$ and $\mathcal{H}=-2.2$ from the strong and electromagnetic decays of the decuplet baryons \cite{butl1992}. The $\sigma_{\pi N}=59.1\pm3.5$ MeV \cite{hofe2015} from Roy–Steiner equations is taken in our calculations. Currently, we do not have enough data to determine all parameters in the chiral limit. Consequently, we can only use their physical values for renormalization, which may present challenges to the stability of the fitting process.

\begin{table*}[!b]
\centering
\begin{threeparttable}
\caption{\label{fittingresult}Values of the LECs for the fits with the phase shifts from the Roy-Steiner equations (RS) and the WI08 Solution. For a detailed description of these fits, see the main text. }
\begin{tabular}{ccccccc}
\midrule
\toprule
 & Fit RS & Fit WI08 &\\
\midrule
$b_0$ ($\text{GeV}^{-1}$)&$-1.32\pm 0.03$&$-1.17\pm 0.03$&\\
\midrule
$b_D$ ($\text{GeV}^{-1}$)&$0.13\pm 0.00$&$0.13\pm 0.00$&\\
\midrule
$b_F$ ($\text{GeV}^{-1}$)&$-0.59\pm 0.02$&$-0.60\pm 0.02$&\\
\midrule
$C_1$ ($\text{GeV}^{-1}$)&$-6.95\pm 0.11$&$-3.97\pm 0.07$&\\
\midrule
$C_2$ ($\text{GeV}^{-1}$)&$4.48\pm 0.13$&$1.75\pm 0.04$&\\
\midrule
$C_3$ ($\text{GeV}^{-1}$)&$0.87\pm 0.02$&$0.09\pm 0.00$&\\
\midrule
$H_1$ ($\text{GeV}^{-2}$)&$2.69\pm 0.08$&$-1.36\pm 0.04$&\\
\midrule
$H_2$ ($\text{GeV}^{-2}$)&$-2.77\pm 0.08$&$-5.43\pm 0.14$&\\
\midrule
$H_3$ ($\text{GeV}^{-2}$)&$5.77\pm 0.12$&$10.38\pm 0.24$&\\
\midrule
$H_4$ ($\text{GeV}^{-2}$)&$18.38\pm 0.48$&$17.53\pm 0.39$&\\
\midrule
$C^T$ ($\text{GeV}^{-1}$)&$-7.03\pm 0.19$&$-3.99\pm 0.09$&\\
\midrule
$H_1^T$ ($\text{GeV}^{-2}$)&$-104.93\pm 2.66$&$-82.57\pm 2.15$&\\
\midrule
$H_2^T$ ($\text{GeV}^{-2}$)&$12.41\pm 0.35$&$-30.13\pm 0.87$&\\
\midrule
$M_0$ ($\text{MeV}$)&$559.91\pm 32.56$&$714.11\pm 34.60$&\\
\midrule
$\chi^2/\text{d.o.f.}$&$0.98$&$3.45$&\\
\bottomrule
\midrule
\end{tabular}
\end{threeparttable}
\end{table*}

There are various strategies to determine the constants. We proceed here following the similar approach as in refs.~\cite{huan20201, huan20202}, that is, we fit to the phase shifts provided by various partial wave analyses. Furthermore, the octect-baryon masses ($M_{N,\Sigma,\Xi,\Lambda}$) and the $\sigma_{\pi N}$ term are used simultaneously in our fits. As input we use the phase shifts in the range of 1125-1220 MeV CMS total energy from the results of the Roy–Steiner  (RS) equations \cite{hofe2016} and the WI08 Solution of the SAID-online \cite{work2012,SAID}. Since no uncertainties are available for the WI08 solution, we give $4\%$ for all partial waves.

The resulting $M_0$ and LECs can be found in Table~\ref{fittingresult}. The uncertainty for the respective parameter is statistical, and it measures how much a particular
parameter can be changed while maintaining a good description of the fitting data. Nevertheless, the parameters cannot really vary independently of each other because of the mutual correlations, as detailed in refs.~\cite{doba2014,carl2016}. Most of the LECs are of natural size, i.e. of order one. However, the LEC combinations from $\mathcal{O}(\epsilon^3)$ are large. This can be the result of significant cancellations between the loop graphs with intermediate decuplets and the counterterms. Similarly, in the third order SSE calculation of $\pi N$ scattering within the SU(2) case, this phenomenon was also observed~\cite{fett2001}. For the RS data, the $\chi^2/\text{d.o.f.}$ is near one, that means we obtain a very good fit. For the WI08 data, the $\chi^2/\text{d.o.f.}$ is larger than one. The reason for this slightly large $\chi^2/\text{d.o.f.}$ is that the phase shifts from WI08 Solution are not sufficiently accurate.

\begin{figure}[p]
\centering
\includegraphics[height=21.4cm,width=13.7cm]{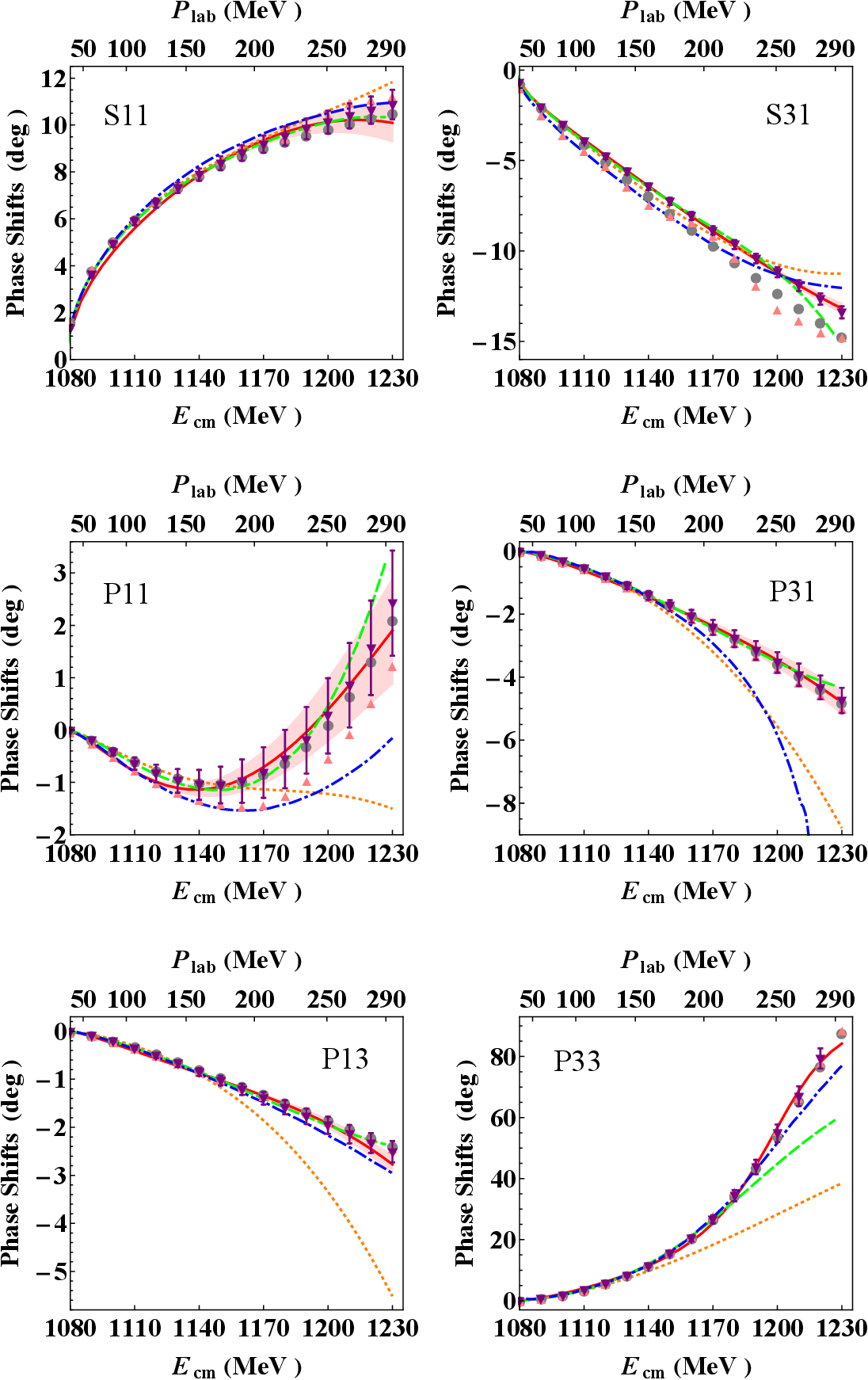}% Here is how to import EPS art
\caption{\label{fig:RS}Results of the $\pi N$ phase shifts versus the CMS total energy $E_{\text{cm}}$ or the pion laboratory momentum $|\bm{p}_{\text{lab}}|$ in pion-nucleon ($\pi N$) scattering. The phase shifts for the fits are from the Roy-Steiner equations.  The red solid lines refer to our results [$\mathcal{O}(\epsilon^3)$] for pion-nucleon phase shifts. The orange dotted lines present the third order [$\mathcal{O}(p^3)$] results without decuplet \cite{huan20201}. The green dashed lines present the fourth order [$\mathcal{O}(p^4)$] results without decuplet \cite{huan20202}. The blue dash-dotted lines present the SU(2) results from ref.~\cite{fett2001}. The purple lower triangles with error bars denote the Roy-Steiner-equation results from ref.~\cite{hofe2016}. The gray circles and pink triangles denote the WI08 and KA84 solutions from SAID \cite{SAID}, respectively. The error bands for our results are generated by the standard error propagation formula.}
\end{figure}

\begin{figure}[p]
\centering
\includegraphics[height=21.4cm,width=13.7cm]{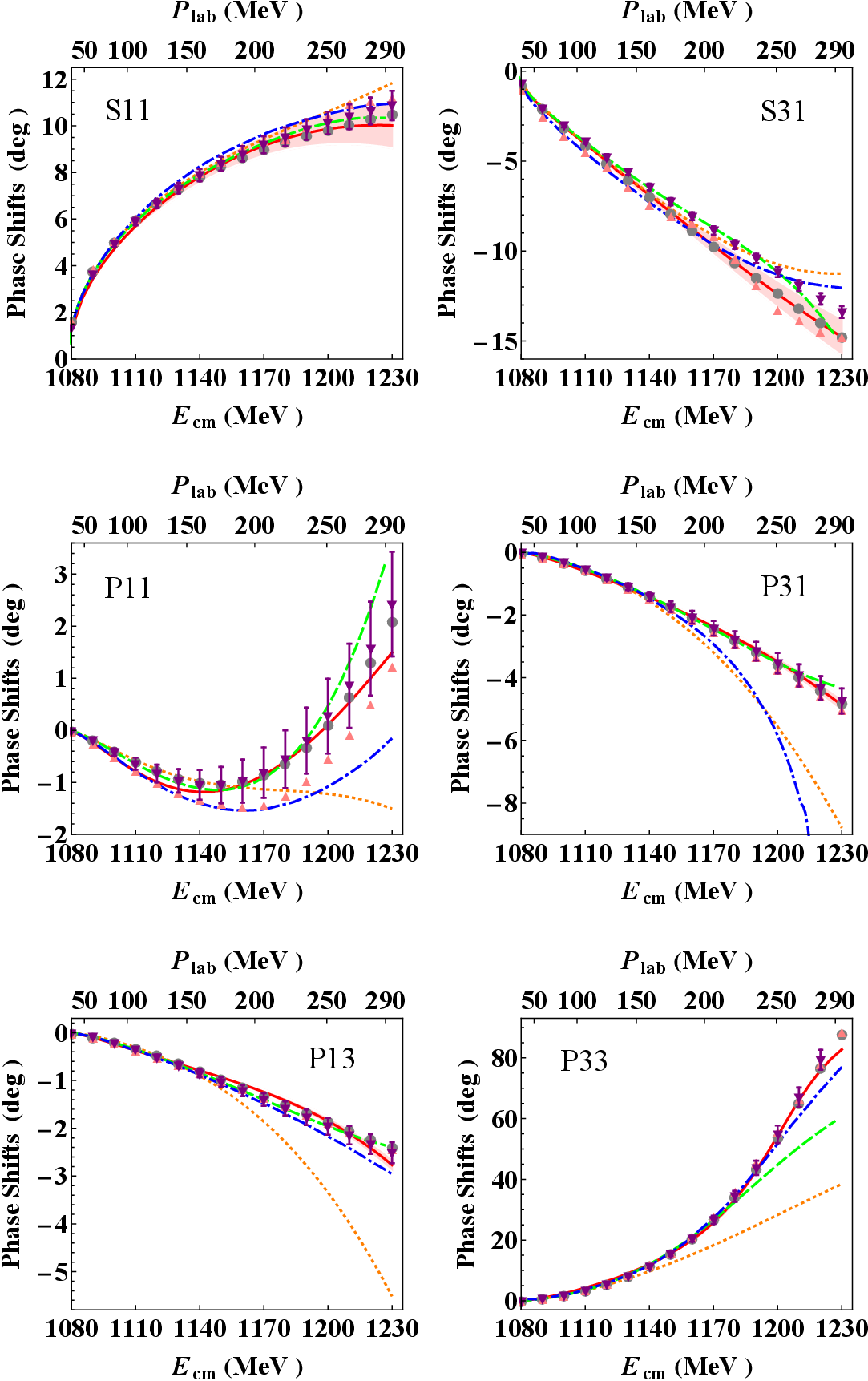}% Here is how to import EPS art
\caption{\label{fig:WI08}Results of the $\pi N$ phase shifts versus the CMS total energy $E_{\text{cm}}$ or the pion laboratory momentum $|\bm{p}_{\text{lab}}|$ in pion-nucleon ($\pi N$) scattering. The phase shifts for the fits are from the WI08 Solution. The notation is the same as in Fig.~\ref{fig:RS}.}
\end{figure}

The phase shifts from the resulting fits and predictions based on the RS and WI08 are presented in Figs.~\ref{fig:RS} and \ref{fig:WI08}, respectively. These results are compared with those obtained from the SU(2) framework and with results derived from third and fourth-order chiral amplitudes without decuplet contribution. The partial waves are denoted by $L_{2I,2J}$ with $L$ the angular momentum, $I$ the total isospin, and $J$ the total angular momentum. Our results [$\mathcal{O}(\epsilon^3)$] for pion-nucleon phase shifts are clearly better than the ones based on the third order [$\mathcal{O}(p^3)$] results without decuplet \cite{huan20201} and comparable to the fourth order [$\mathcal{O}(p^4)$] results without decuplet \cite{huan20202}, although the overall description is still slightly better in the latter case. As expected, the most prominent improvement can be found in the $P_{33}$ partial wave, which is well described up to the $\Delta$ pole. Now, we compare our results with those from the SU(2) framework. For all partial waves, except for the $P_{11}$ and $P_{31}$ waves, we can believe our results to be consistent with those of SU(2) within the fitting accuracy and error range. However, for the $P_{11}$ and $P_{31}$ waves, our results are considerably better than those of the SU(2) framework. This suggests that the inclusion of strange degrees of freedom in the amplitudes has enabled us to achieve a good fit at this order. It is essential to recognize that this does not imply the SU(3) framework is better than the SU(2) framework, as the calculations up to the third order do not provide fully convergent results, as discussed below. Clearly, our results for the two fits are in good agreement with the phase shifts from RS and WI08, respectively. Consequently, the slightly large $\chi^2/\text{d.o.f.}$ for WI08 suggests that the description of the baryon masses or the $\sigma_{\pi N}$ may not be entirely accurate.

The baryon masses and the $\sigma_{\pi N}$ of the resulting fits are shown in Table~\ref{baryonmasses}. We can see that the baryon masses for the two fits are in good agreement with the data from PDG. The value of the $\sigma_{\pi N}$ from the RS fit is consistent with the value of the input within errors. It is not unexpected because the value of the input is from Roy-Steiner equation. Meanwhile, the value of the $\sigma_{\pi N}$ from WI08 fit is much smaller than the value of the input, which can lead to the slightly large $\chi^2/\text{d.o.f.}$ for the WI08 fit. Above all, we obtain a reliable fit by using accurate phase shifts and $\sigma_{\pi N}$ from RS equations in combination with the baryon masses.

\begin{table*}[!b]
\centering
\begin{threeparttable}
\caption{\label{baryonmasses}Results of the baryon masses and the $\sigma_{\pi N}$ the fits with the physical baryon masses, the $\sigma_{\pi N}$ and the $\pi N$ phase shifts simultaneously. The two KN $\sigma$-terms and the $\sigma_{s N}$  are predicted by using the resulting constants of the fits. Note that the value marked with an asterisk is unphysical.}
\begin{tabular}{ccccccc}
\midrule
\toprule
&Input & Fit RS & Fit WI08 & \\
\midrule
$M_N$ (MeV)&$938.92\pm 1.29$&$939.08\pm 8.51$& $939.06\pm 5.86$& \\
\midrule
$M_\Sigma$ (MeV)&$1191.01\pm 4.86$&$1190.37\pm 12.90$& $1191.28\pm 14.19$& \\
\midrule
$M_\Xi$ (MeV)&$1318.26\pm 6.30$&$1322.83\pm 25.05$& $1323.88\pm 27.22$& \\
\midrule
$M_\Lambda$ (MeV)&$1115.68\pm 5.58$&$1111.25\pm 13.32$& $1111.63\pm 13.06$& \\
\midrule
$\sigma_{\pi N}$ (MeV)&$59.1\pm3.5$  &$52.08\pm 2.61$&$40.30\pm 2.31$&  \\
\midrule
$\sigma_{K N}^{(1)}$ (MeV)&—/— &$375.07\pm 33.02$&$226.48\pm 29.93$&\\
\midrule
$\sigma_{K N}^{(2)}$ (MeV)&—/— &$275.32\pm 32.24$&$126.59\pm 35.62$&\\
\midrule
$\sigma_{s N}$ (MeV)&—/— &$37.99\pm 31.11$&$-105.38\pm 35.52^{\ast}$&\\
\bottomrule
\midrule
\end{tabular}
\end{threeparttable}
\end{table*}

We can predict the KN $\sigma$-terms and the $\sigma_{sN}$ by using the resulting constants of the two fits. The values of the KN $\sigma$-terms and $\sigma_{sN}$ can be found in Table~\ref{baryonmasses}. Obviously, the results from RS are larger than the ones from WI08. This is not surprising because the value of the $\sigma_{\pi N}$ from RS is also larger than the one from WI08. The values for $\sigma$-terms from WI08 are consistent with those in ref.~\cite{bern19931}. This means that the KN $\sigma$-terms are proportional to the $\sigma_{\pi N}$. We obtain a very small value for $\sigma_{sN}$ in the RS fit, significantly smaller than the value reported in ref.~\cite{Ellis:2018dmb}. However, the lattice QCD result is 43(8) MeV \cite{Lin:2011ab}, which is consistent with our findings within errors. Additionally, a small strangeness content in the proton was found in ref.~\cite{Severt:2019sbz}. In the WI08 fit, we find a negative value for $\sigma_{sN}$, which is clearly unphysical and unreasonable. This issue arises because $\sigma_{sN}$ is highly sensitive to the LECs ($b_{D,F,0}$), and the inaccurate phase shifts from WI08 prevent $b_{D,F,0}$ from properly describing $\sigma_{sN}$. In conclusion, we achieve reasonable KN $\sigma$ terms using data from the RS equations as inputs.

Let us apply the above constants (Table~\ref{fittingresult}) to estimate the pion-nucleon scattering lengths and scattering volumes. The scattering lengths and scattering volumes are determined by utilizing an incident pion momentum of 1 MeV and approximating their values at the threshold. We provide the values of the scattering lengths and scattering volumes in Table~\ref{pinthresholdparameters}, alongside comparisons to the values from various analyses. The errors associated with our results are also statistical and can be calculated using the standard error propagation formula from the fitting constants. First, we observe that our results of the RS and WI08 for both scattering lengths and scattering volumes are consistent within errors. There are two experimental values for scattering lengths in Table ~\ref{pinthresholdparameters}. The former, EXP2015, are obtained by combining with the analysis of the results from refs.~\cite{baru2011,baru20112,hofe2013,henn2014}, as done in ref.~\cite{hofe2015}. Our results for scattering lengths are consistent with those values within errors. At present, there are no accurate experimental values for scattering volumes in comparison to our results. However, our predictions for scattering lengths and scattering volumes are expected to be reliable.

\begin{table*}[!t]
\centering
\begin{threeparttable}
\caption{\label{pinthresholdparameters}
Values of the $S$- and $P$-wave scattering lengths and scattering volumes. The errors for our results are derived using the standard error propagation formula based on the fitting constants, and they reflect only statistical uncertainties.}
\begin{tabular}{cccccccccccc}
\midrule
\toprule
 &Fit RS &Fit WI08 & EXP2015 \cite{hofe2015}& EXP2001 \cite{schr2001}  &  \\
\midrule
$a_{0+}^{3/2}$ (fm) & $-0.119\pm0.004$ & $-0.121\pm0.005$ &$-0.122\pm 0.003$ &$-0.121 \pm 0.003 $&   \\
\midrule
$a_{0+}^{1/2}$ (fm)  & $0.214\pm0.003$ & $0.223\pm0.006$& $0.240\pm 0.003$ & $0.250^{+0.006}_{-0.004}$&  \\
\midrule
 $a_{1+}^{3/2}$ ($\text{fm}^3$) & $0.853\pm0.008$& $0.871\pm0.015$ &—/—  & —/—  & \\
\midrule
 $a_{1+}^{1/2}$ ($\text{fm}^3$)  & $-0.115\pm0.003$ & $-0.104\pm0.001$&—/—  &—/— &  \\
  \midrule
 $a_{1-}^{3/2}$ ($\text{fm}^3$) & $-0.156\pm0.005$ & $-0.150\pm0.002$&—/—  &—/— &   \\
\midrule
 $a_{1-}^{1/2}$ ($\text{fm}^3$) & $-0.268\pm0.008$ & $-0.260\pm0.003$& —/—  &—/— &   \\
\bottomrule
\midrule
\end{tabular}
\end{threeparttable}
\end{table*}

\begin{table*}[!b]
\centering
\begin{threeparttable}
\caption{\label{scattering length}Convergence of the $S$-wave scattering lengths. $\mathcal{O}(\epsilon^n)$ means that all terms up-to-and-including order $n$ are given. The errors for our results are derived using the standard error propagation formula based on the fitting constants, and they reflect only statistical uncertainties.}
\begin{tabular}{cccccccccccc}
\midrule
\toprule
 &  &  $\mathcal{O}(\epsilon)$ & $\mathcal{O}(\epsilon^2)$  & $\mathcal{O}(\epsilon^3)$ &\\
\midrule
$a_{0+}^+$ (fm) & \thead{RS \\ WI08}  &$ \thead{0.000 \\ 0.000} $ &$\thead{-0.006\pm0.003 \\ -0.011\pm0.006}$&$\thead{-0.008\pm0.003 \\ -0.006\pm0.004} $&   \\
\midrule
$a_{0+}^{-}$ (fm)  & \thead{RS \\ WI08} & $\thead{0.113 \\ 0.113}$ & $\thead{0.113\pm0.000 \\ 0.113\pm0.000} $ & $\thead{0.111\pm0.002 \\ 0.115\pm0.003}$&  \\
\bottomrule
\midrule
\end{tabular}
\end{threeparttable}
\end{table*}

\begin{figure}[!t]
\centering
\includegraphics[height=21.4cm,width=13.7cm]{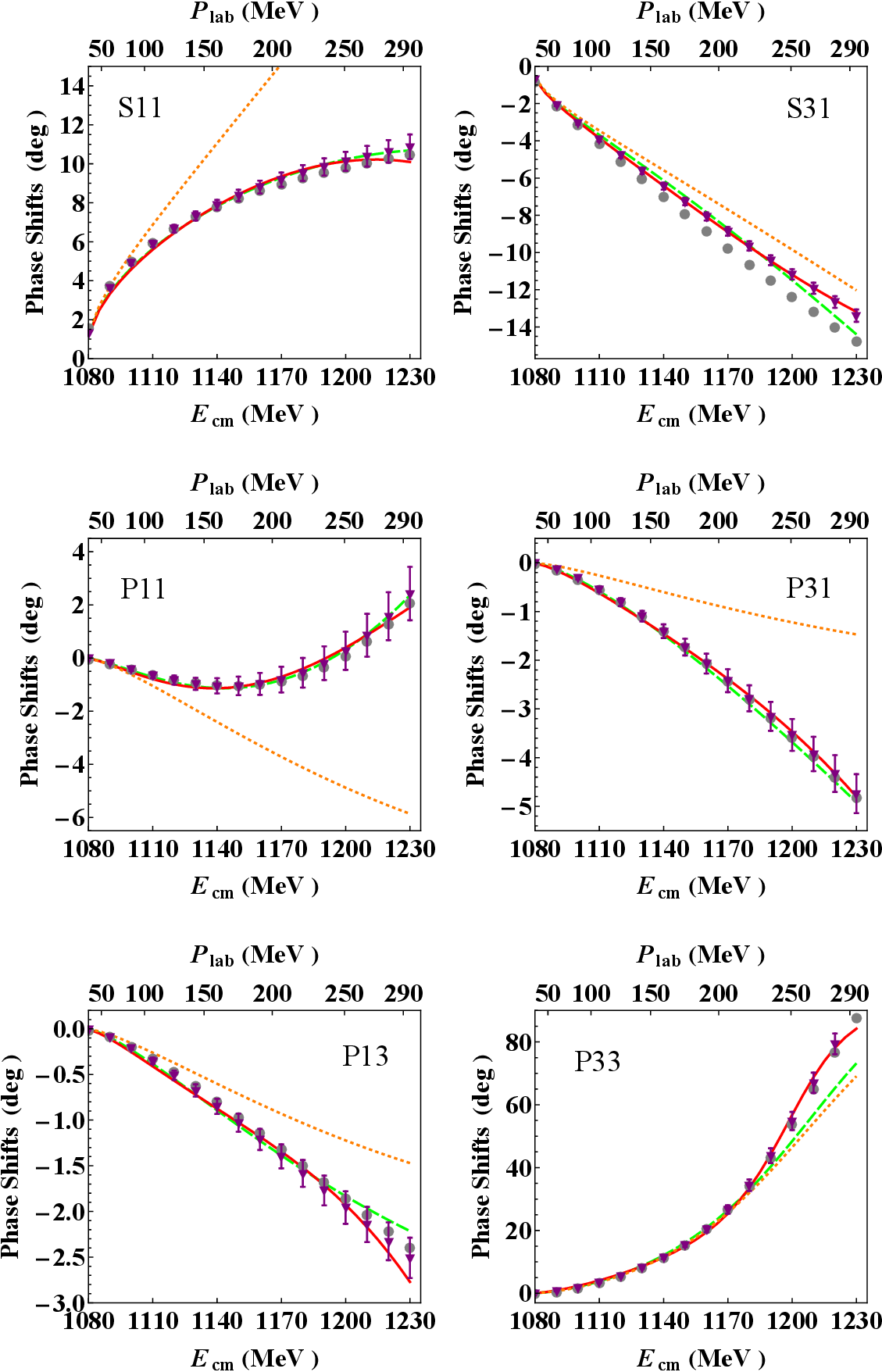}% Here is how to import EPS art
\caption{\label{fig:refitRS}Fits and predictions for the RS phase shifts as a function of the CMS total energy $E_{\text{cm}}$ or the pion laboratory momentum $|\bm{p}_{\text{lab}}|$ to first (orange-dotted lines), second (green-dashed lines) and third (red-solid lines) order in the small scale expansion. The purple lower triangles with error bars denote the Roy-Steiner equation results from ref.~\cite{hofe2016} and the gray circles denote the WI08 solutions from SAID \cite{SAID}.}
\end{figure}

\begin{figure}[!t]
\centering
\includegraphics[height=21.4cm,width=13.7cm]{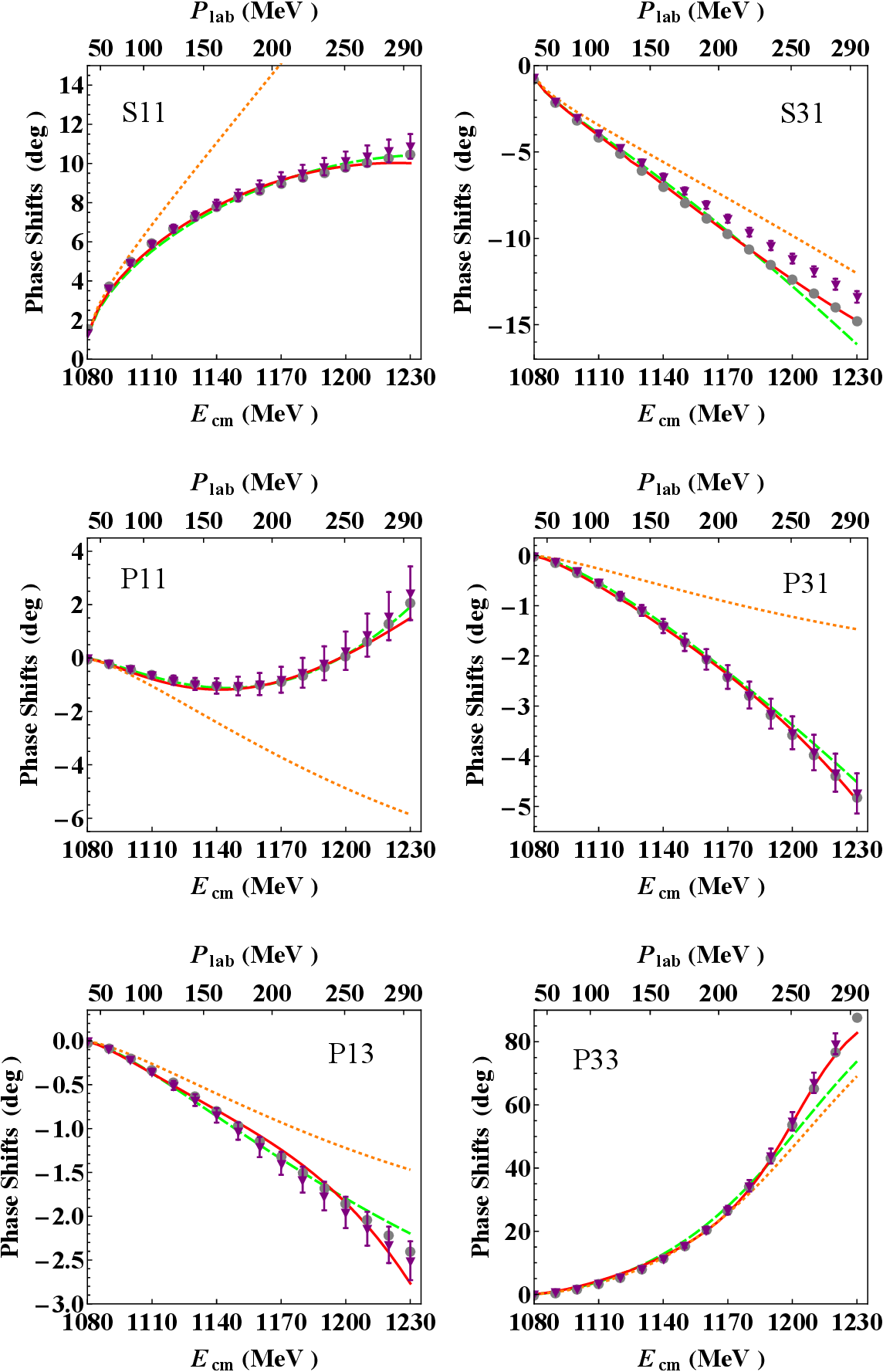}% Here is how to import EPS art
\caption{\label{fig:refitWI08}Fits and predictions for the WI08 phase shifts as a function of the CMS total energy $E_{\text{cm}}$ or the pion laboratory momentum $|\bm{p}_{\text{lab}}|$ to first (orange-dotted lines), second (green-dashed lines) and third (red-solid lines) order in the small scale expansion. The purple lower triangles with error bars denote the Roy-Steiner equation results from ref.~\cite{hofe2016} and the gray circles denote the WI08 solutions from SAID \cite{SAID}.}
\end{figure}

It is essential to discuss the convergence issue. For this, we redo the fits for amplitudes at both first and second orders in the small scale expansion. At the leading order, the amplitude is characterized solely by the coupling constant $\mathcal{C}$, whereas the second order introduces seven LECs along with the baryon mass $M_0$ in the chiral limit. The resulting phase shifts are shown in Fig.~\ref{fig:refitRS} for the RS analysis and in Fig.~\ref{fig:refitWI08} for the WI08 case. While the first-order results adequately describe the $P_{33}$ partial wave, the second-order fits demonstrate a quality comparable to the third-order ones, albeit with a significantly improved $\chi^2/\text{d.o.f.}$ at the third order. This trend aligns with expectations based on the predominant role of Born graphs incorporating the decuplet, an insight gained from various models including the decuplet and explicit calculations of Compton scattering off nucleons within the SSE framework, as detailed in ref.~\cite{hemm1998}. Contrastingly, in the chiral expansion, the transition from second to third order involves substantial corrections, while the progression from third to fourth order typically introduces only modest changes (refer to refs.~\cite{huan20201,huan20202}). This indicates that in channels where delta resonances are significant, resumming higher-order terms in the chiral expansion is crucial.

\begin{table*}[!t]
\centering
\begin{threeparttable}
\caption{\label{sigmaterms}Convergence of the $\sigma$ terms from the RS fit. $\mathcal{O}(\epsilon^n)$ means that all terms up-to-and-including order $n$ are given. The errors for our results are derived using the standard error propagation formula based on the fitting constants, and they reflect only statistical uncertainties. }
\begin{tabular}{cccccccccccc}
\midrule
\toprule
 &   $\mathcal{O}(\epsilon^2)$  & $\mathcal{O}(\epsilon^3)$ &\\
\midrule
$\sigma_{\pi N}$ (MeV) &$62.58\pm 5.24 $&$52.08 \pm 2.61$&   \\
\midrule
$\sigma_{KN}^{(1)}$ (MeV) & $671.98\pm 53.29$ & $375.07\pm 33.02 $&  \\
\midrule
$\sigma_{KN}^{(2)}$ (MeV) & $528.35 \pm 46.34  $ & $ 275.32\pm 32.24 $&  \\
\midrule
$\sigma_{sN}$ (MeV) & $448.83 \pm 52.23 $ & $ 37.99 \pm 31.11 $&  \\
\bottomrule
\midrule
\end{tabular}
\end{threeparttable}
\end{table*}

It is also interesting to investigate the convergence of the S–wave scattering lengths, as it has been done for the chiral expansion in ref.~\cite{huan20201}. In the small scale expansion the results are collected in Table~\ref{scattering length}. The convergence for the isovector scattering length is similar to what is obtained in the SU(2) case~\cite{fett2001}. The isoscalar S–wave scattering length receives a minor correction when going from second to third order, which is different from the SU(2) case. Obviously, our results in the SU(3) case have a much improved discription of scattering lengths.

To examine the accuracy of our predicted KN $\sigma$ terms, we investigate the convergence of the $\sigma$ terms. In the small scale expansion the results can be found in Table~\ref{sigmaterms}. We observe that the results at the order $\mathcal{O}(\epsilon)$ are all zero, with results starting from the $\mathcal{O}(\epsilon^2)$ order. At $\mathcal{O}(\epsilon^2)$, we obtain reasonable values for $\sigma_{\pi N}$, but we also find significantly large values for the KN $\sigma$ terms and $\sigma_{sN}$. At $\mathcal{O}(\epsilon^3)$, the $\sigma_{\pi N}$ value is somewhat smaller, but it remains consistent with the $\mathcal{O}(\epsilon^2)$ results within the error range, while the KN $\sigma$ terms and $\sigma_{sN}$ values are much smaller. Clearly, we have achieved good convergence for $\sigma_{\pi N}$, but not for the KN $\sigma$ terms and $\sigma_{sN}$. This indicates that our predicted values for KN $\sigma$ terms contain a non-negligible systematic error due to the truncation at higher orders.

In summary, we calculated the complete T matrices with decuplet contributions for pion-nucleon scattering to order $\mathcal{O}(\epsilon^3)$ in SU(3) HB$\chi$PT. We simultaneously fitted the empirical phase shifts of $\pi$N scattering, the experimental octet-baryon masses and the value of $\sigma_{\pi N}$ to determine the $M_0$ and the LECs. This led to an excellent description of the phase shifts for all partial waves below 1230 MeV total energy in the CMS. The $M_0$ and LEC uncertainties were analyzed through statistical regression. We obtained the reasonable KN $\sigma$ terms ($\sigma_{KN}^{(1)}=375.07\pm33.02$ MeV, $\sigma_{KN}^{(2)}=275.32\pm32.24$ MeV) using the data from RS equations as inputs. We estimated the scattering lengths and scattering volumes using the constants from the results of fits. The scattering lengths turned out to be in good agreement with those of available experimental data. Finally, we discussed the convergence of the phase shifts and scattering lengths for $\pi N$ scattering. The results demonstrated a significant improvement in convergence, especially in the $P_{33}$ channel. The analysis of the convergence of $\sigma$ terms implies that our predicted values for KN $\sigma$ terms contain a non-negligible systematic error due to truncation at higher orders. The inability to achieve good convergence may lead to the instability of the fits when using physical values for the renormalization of meson decay constants, mass differences of octets and decuplets, and other related quantities.

\section*{Acknowledgments}
This work is supported by the National Natural Science Foundation of
China under Grants No. 12075126, and No. 12147127. B.-L. Huang is also supported by the Inner Mongolia Autonomous Region Natural Science Fund (No. 2024MS01016), the Research Support for High-Level Talent at the Autonomous Region Level (No. 13100-15112042) and the Junma Program High-Level Talent Reseach Start-up Fund (No. 10000-23112101/085). We thank Norbert Kaiser (Technische Universit\"{a}t München) and Shi-Lin Zhu (Peking University) for very helpful discussions.

\clearpage
\appendix\markboth{Appendix}{Appendix}
\renewcommand{\thesection}{\Alph{section}}
\numberwithin{equation}{section}
\section{One-loop amplitudes with decuplet contribution}
\label{oneloopamplitudes}
In this appendix, we present the amplitudes from nonvanishing one-loop diagrams with decuplet contribution. The one-loop chiral corrections are given by\\

Figs. (1)+(2):
\begin{align}
\label{A1}
V_{(\pi N)}^{(3/2)}=&\frac{\mathcal{C}^2}{18f^4(\delta-w)w}(D+F)(-2m_\pi^2+t+2w^2)\Big[(3D+F)J_2(m_K,-\delta)\nonumber\\&-(3D+F)J_2(m_K,-w)+8(D+F)\Big(J_2(m_\pi,-\delta)-J_2(m_\pi,-w)\Big)\Big],
\end{align}
\begin{align}
\label{A2}
W_{(\pi N)}^{(3/2)}=&\frac{\mathcal{C}^2}{9f^4(w-\delta)w}(D+F)\Big[(3D+F)J_2(m_K,-\delta)-(3D+F)J_2(m_K,-w)\nonumber\\&+8(D+F)\Big(J_2(m_\pi,-\delta)-J_2(m_\pi,-w)\Big)\Big],
\end{align}
\begin{align}
\label{A3}
V_{(\pi N)}^{(1/2)}=&\frac{\mathcal{C}^2}{36f^4w(w^2-\delta^2)}(D+F)(-2m_\pi^2+t+2w^2)\Bigg\{\delta\Big[-3(3D+F)J_2(m_K,w)\nonumber\\&+8(D+F)\Big(4J_2(m_\pi,-\delta)-J_2(m_\pi,-w)-3J_2(m_\pi,w)\Big)\Big]\nonumber\\&+2(3D+F)J_2(m_K,-\delta)(2\delta-w)+\Big[3(3D+F)J_2(m_K,w)\nonumber\\&-8(D+F)\Big(2J_2(m_\pi,-\delta)+J_2(m_\pi,-w)-3J_2(m_\pi,w)\Big)\Big]w\nonumber\\&-(3D+F)J_2(m_K,-w)(\delta+w)\Bigg\},
\end{align}
\begin{align}
\label{A4}
W_{(\pi N)}^{(1/2)}=&\frac{\mathcal{C}^2}{18f^4w}(D+F)\Bigg\{\frac{1}{\delta-w}\Big[(3D+F)J_2(m_K,-\delta)-(3D+F)J_2(m_K,-w)\nonumber\\&+8(D+F)\Big(J_2(m_\pi,-\delta)-J_2(m_\pi,-w)\Big)\Big]+\frac{3}{\delta+w}\Big[-(3D+F)J_2(m_K,-\delta)\nonumber\\&+(3D+F)J_2(m_K,w)-8(D+F)\Big(J_2(m_\pi,-\delta)-J_2(m_\pi,w)\Big)\Big]\Bigg\}.
\end{align}
Figs. (3)+(4):
\begin{align}
\label{A5}
V_{(\pi N)}^{(3/2)}=&-\frac{\mathcal{C}^2}{18wf^4}(D+F)(-2m_\pi^2+t+2w^2)\Big[(3D+F)\Gamma_2(m_K,-\delta-w)\nonumber\\&+8(D+F)\Gamma_2(m_\pi,-\delta-w)\Big],
\end{align}
\begin{align}
\label{A6}
W_{(\pi N)}^{(3/2)}=&\frac{\mathcal{C}^2}{9wf^4}(D+F)\Big[(3D+F)\Gamma_2(m_K,-\delta-w)+8(D+F)\Gamma_2(m_\pi,-\delta-w)\Big],
\end{align}
\begin{align}
\label{A7}
V_{(\pi N)}^{(1/2)}=&\frac{\mathcal{C}^2}{36wf^4}(D+F)(-2m_\pi^2+t+2w^2)\Big[(3D+F)\Gamma_2(m_K,-\delta-w)\nonumber\\&+3(3D+F)\Gamma_2(m_K,-\delta+w)+8(D+F)\Big(\Gamma_2(m_\pi,-\delta-w)\nonumber\\&+3\Gamma_2(m_\pi,-\delta+w)\Big)\Big],
\end{align}
\begin{align}
\label{A8}
W_{(\pi N)}^{(1/2)}=&-\frac{\mathcal{C}^2}{18wf^4}(D+F)\Big[(3D+F)\Gamma_2(m_K,-\delta-w)-3(3D+F)\Gamma_2(m_K,-\delta+w)\nonumber\\&+8(D+F)\Big(\Gamma_2(m_\pi,-\delta-w)-3\Gamma_2(m_\pi,-\delta+w)\Big)\Big].
\end{align}
Figs. (5)+(6):
\begin{align}
\label{A9}
V_{(\pi N)}^{(3/2)}=&\frac{2\mathcal{C}^2}{27f^4(\delta^2-w^2)}(-m_\pi^2+\frac{t}{2}+w^2)\Big[-(D^2+6DF-3F^2)\Gamma_2(m_K,-w)(\delta-w)\nonumber\\&-3(D^2+6DF-3F^2)\Gamma_2(m_K,w)(\delta+w)+3(D+F)^2\Big(\Gamma_2(m_\pi,-w)(-\delta+w)\nonumber\\&-3\Gamma_2(m_\pi,w)(\delta+w)\Big)\Big],
\end{align}
\begin{align}
\label{A10}
W_{(\pi N)}^{(3/2)}=&\frac{\mathcal{C}^2}{27f^4}\Big\{\frac{3}{\delta-w}(D^2+6DF-3F^2)\Gamma_2(m_K,w)+\frac{9}{\delta-w}(D+F)^2\Gamma_2(m_\pi,w)\nonumber\\&-\frac{1}{\delta+w}\Big[(D^2+6DF-3F^2)\Gamma_2(m_K,-w)+3(D+F)^2\Gamma_2(m_\pi,-w)\Big]\Big\},
\end{align}
\begin{align}
\label{A11}
V_{(\pi N)}^{(1/2)}=&-\frac{4\mathcal{C}^2}{27f^4(\delta+w)}(-2m_\pi^2+t+2w^2)\Big[(D^2+6DF-3F^2)\Gamma_2(m_K,-w)\nonumber\\&+3(D+F)^2\Gamma_2(m_\pi,-w)\Big],
\end{align}
\begin{align}
\label{A12}
W_{(\pi N)}^{(1/2)}=&-\frac{4\mathcal{C}^2}{27f^4(\delta+w)}\Big[(D^2+6DF-3F^2)\Gamma_2(m_K,-w)+3(D+F)^2\Gamma_2(m_\pi,-w)\Big].
\end{align}
Figs. (7)+(8):
\begin{align}
\label{A13}
V_{(\pi N)}^{(3/2)}=&\frac{5\mathcal{C}^2\mathcal{H}}{81f^4w^2}(D+F)(-2m_\pi^2+t+2w^2)\Big[J_2(m_K,-\delta)-J_2(m_K,-\delta-w)\nonumber\\&+5J_2(m_\pi,-\delta)-5J_2(m_\pi,-\delta-w)\Big],
\end{align}
\begin{align}
\label{A14}
W_{(\pi N)}^{(3/2)}=&-\frac{10\mathcal{C}^2\mathcal{H}}{81f^4w^2}(D+F)\Big[J_2(m_K,-\delta)-J_2(m_K,-\delta-w)\nonumber\\&+5J_2(m_\pi,-\delta)-5J_2(m_\pi,-\delta-w)\Big],
\end{align}
\begin{align}
\label{A15}
V_{(\pi N)}^{(1/2)}=&\frac{5\mathcal{C}^2\mathcal{H}}{162f^4w^2}(D+F)(-2m_\pi^2+t+2w^2)\Big[2J_2(m_K,-\delta)+J_2(m_K,-\delta-w)\nonumber\\&-3J_2(m_K,-\delta+w)+5\Big(2J_2(m_\pi,-\delta)+J_2(m_\pi,-\delta-w)\nonumber\\&-3J_2(m_\pi,-\delta+w)\Big)\Big],
\end{align}
\begin{align}
\label{A16}
W_{(\pi N)}^{(1/2)}=&\frac{5\mathcal{C}^2\mathcal{H}}{81f^4w^2}(D+F)\Big[4J_2(m_K,-\delta)-J_2(m_K,-\delta-w)-3J_2(m_K,-\delta+w)\nonumber\\&+20J_2(m_\pi,-\delta)-5J_2(m_\pi,-\delta-w)-15J_2(m_\pi,-\delta+w)\Big].
\end{align}
Figs. (9)+(10):
\begin{align}
\label{A17}
V_{(\pi N)}^{(3/2)}=&\frac{\mathcal{C}^4}{324f^4(\delta^2-w^2)}(-2m_\pi^2+t+2w^2)\Big[4J_2(m_K,-\delta)-J_2(m_K,-w)\nonumber\\&-3J_2(m_K,w)+8J_2(m_\pi,-\delta)-2J_2(m_\pi,-w)-6J_2(m_\pi,w)\Big],
\end{align}
\begin{align}
\label{A18}
W_{(\pi N)}^{(3/2)}=&-\frac{\mathcal{C}^4}{324f^4(\delta^2-w^2)}\Big[2J_2(m_K,-\delta)+J_2(m_K,-w)-3J_2(m_K,w)\nonumber\\&+4J_2(m_\pi,-\delta)+2J_2(m_\pi,-w)-6J_2(m_\pi,w)\Big],
\end{align}
\begin{align}
\label{A19}
V_{(\pi N)}^{(1/2)}=&\frac{\mathcal{C}^4}{81f^4(\delta^2-w^2)}(-2m_\pi^2+t+2w^2)\Big[J_2(m_K,-\delta)-J_2(m_K,-w)\nonumber\\&+2J_2(m_\pi,-\delta)-2J_2(m_\pi,-w)\Big],
\end{align}
\begin{align}
\label{A20}
W_{(\pi N)}^{(1/2)}=&\frac{\mathcal{C}^4}{81f^4(\delta^2-w^2)}\Big[J_2(m_K,-\delta)-J_2(m_K,-w)+2J_2(m_\pi,-\delta)-2J_2(m_\pi,-w)\Big].
\end{align}

Figs. (11)+(12):
\begin{align}
\label{A21}
V_{(\pi N)}^{(3/2)}=&-\frac{5\mathcal{C}^2\mathcal{H}}{324f^4(\delta^2-w^2)}(-2m_\pi^2+t+2w^2)\Bigg\{\delta\Big[-4F\Gamma_2(m_K,-\delta-w)\nonumber\\&-12F\Gamma_2(m_K,-\delta+w)-5(D+F)\Big(\Gamma_2(m_\pi,-\delta-w)+3\Gamma_2(m_\pi,-\delta+w)\Big)\Big]\nonumber\\&+(D-3F)\Gamma_2(m_\eta,-\delta-w)(\delta-w)+\Big[4F\Gamma_2(m_K,-\delta-w)\nonumber\\&-12F\Gamma_2(m_K,-\delta+w)+5(D+F)\Big(\Gamma_2(m_\pi,-\delta-w)-3\Gamma_2(m_\pi,-\delta+w)\Big)\Big]w\nonumber\\&+3(D-3F)\Gamma_2(m_\eta,-\delta+w)(\delta+w)\Bigg\},
\end{align}
\begin{align}
\label{A22}
W_{(\pi N)}^{(3/2)}=&\frac{5\mathcal{C}^2\mathcal{H}}{324f^4(\delta^2-w^2)}\Bigg\{\delta\Big[4F\Gamma_2(m_K,-\delta-w)-12F\Gamma_2(m_K,-\delta+w)\nonumber\\&+5(D+F)\Big(\Gamma_2(m_\pi,-\delta-w)-3\Gamma_2(m_\pi,-\delta+w)\Big)\Big]\nonumber\\&-(D-3F)\Gamma_2(m_\eta,-\delta-w)(\delta-w)-\Big[4F\Gamma_2(m_K,-\delta-w)\nonumber\\&+12F\Gamma_2(m_K,-\delta+w)+5(D+F)\Big(\Gamma_2(m_\pi,-\delta-w)\nonumber\\&+3\Gamma_2(m_\pi,-\delta+w)\Big)\Big]w+3(D-3F)\Gamma_2(m_\eta,-\delta+w)(\delta+w)\Bigg\},
\end{align}
\begin{align}
\label{A23}
V_{(\pi N)}^{(1/2)}=&-\frac{5\mathcal{C}^2\mathcal{H}}{81f^4(\delta+w)}(-2m_\pi^2+t+2w^2)\Big[(D-3F)\Gamma_2(m_\eta,-\delta-w)\nonumber\\&-4F\Gamma_2(m_K,-\delta-w)-5(D+F)\Gamma_2(m_\pi,-\delta-w)\Big],
\end{align}
\begin{align}
\label{A24}
W_{(\pi N)}^{(1/2)}=&\frac{5\mathcal{C}^2\mathcal{H}}{81f^4(\delta+w)}\Big[-(D-3F)\Gamma_2(m_\eta,-\delta-w)+4F\Gamma_2(m_K,-\delta-w)\nonumber\\&+5(D+F)\Gamma_2(m_\pi,-\delta-w)\Big].
\end{align}

Figs. (13)+(14):
\begin{align}
\label{A25}
V_{(\pi N)}^{(3/2)}=&\frac{10\mathcal{C}^2\mathcal{H}^2}{729wf^4}(-m_\pi^2+\frac{t}{2}+w^2)\Bigg\{\frac{3}{\delta-w}\Big[J_2(m_K,-\delta)-J_2(m_K,-\delta+w)\nonumber\\&+5J_2(m_\pi,-\delta)-5J_2(m_\pi,-\delta+w)\Big]+\frac{1}{\delta+w}\Big[-J_2(m_K,-\delta)\nonumber\\&+J_2(m_K,-\delta-w)-5J_2(m_\pi,-\delta)+5J_2(m_\pi,-\delta-w)\Big]\Bigg\},
\end{align}
\begin{align}
\label{A26}
W_{(\pi N)}^{(3/2)}=&\frac{5\mathcal{C}^2\mathcal{H}^2}{729wf^4(w^2-{\delta^2})}\Bigg\{-\delta\Big[3J_2(m_K,-\delta+w)+5\Big(-4J_2(m_\pi,-\delta)\nonumber\\&+J_2(m_\pi,-\delta-w)+3J_2(m_\pi,-\delta+w)\Big)\Big]+\Big[-3J_2(m_K,-\delta+w)\nonumber\\&+5\Big(2J_2(m_\pi,-\delta)+J_2(m_\pi,-\delta-w)-3J_2(m_\pi,-\delta+w)\Big)\Big]w\nonumber\\&+J_2(m_K,-\delta-w)(-\delta+w)+2J_2(m_K,-\delta)(2\delta+w)\Bigg\},
\end{align}
\begin{align}
\label{A27}
V_{(\pi N)}^{(1/2)}=&-\frac{20\mathcal{C}^2\mathcal{H}^2}{729wf^4(\delta+w)}(-2m_\pi^2+t+2w^2)\Big[J_2(m_K,-\delta)-J_2(m_K,-\delta-w)\nonumber\\&+5J_2(m_\pi,-\delta)-5J_2(m_\pi,-\delta-w)\Big],
\end{align}
\begin{align}
\label{A28}
W_{(\pi N)}^{(1/2)}=&-\frac{20\mathcal{C}^2\mathcal{H}^2}{729wf^4(\delta+w)}\Big[J_2(m_K,-\delta)-J_2(m_K,-\delta-w)\nonumber\\&+5J_2(m_\pi,-\delta)-5J_2(m_\pi,-\delta-w)\Big].
\end{align}
Figs. (15)+(16):
\begin{align}
\label{A29}
V_{(\pi N)}^{(3/2)}=0,
\end{align}
\begin{align}
\label{A30}
W_{(\pi N)}^{(3/2)}=&\frac{\mathcal{C}^2}{54f^4(\delta^2-w^2)}\Bigg\{\delta\Big[(D^2+6DF-3F^2)\Gamma_2(m_K,w)\nonumber\\&+6(D+F)^2\Big(-2\Gamma_2(m_\pi,-\delta)+\Gamma_2(m_\pi,-w)+\Gamma_2(m_\pi,w)\Big)\Big]\nonumber\\&-\Big[(D^2+6DF-3F^2)\Gamma_2(m_K,w)+6(D+F)^2\Big(-\Gamma_2(m_\pi,-w)\nonumber\\&+\Gamma_2(m_\pi,w)\Big)\Big]w+(7D^2-12DF+9F^2)\Gamma_2(m_K,-w)(\delta+w)\nonumber\\&-2\Gamma_2(m_K,-\delta)\Big(\delta(4D^2-3DF+3F^2)+3(D-2F)(D-F)w\Big)\Bigg\},
\end{align}
\begin{align}
\label{A31}
V_{(\pi N)}^{(1/2)}=0,
\end{align}
\begin{align}
\label{A32}
W_{(\pi N)}^{(1/2)}=&-\frac{\mathcal{C}^2}{54f^4(\delta^2-w^2)}\Bigg\{\delta\Big[(11D^2-15DF+12F^2)\Gamma_2(m_K,w)\nonumber\\&+12(D+F)^2\Big(-2\Gamma_2(m_\pi,-\delta)+\Gamma_2(m_\pi,-w)+\Gamma_2(m_\pi,w)\Big)\Big]\nonumber\\&+\Big[(-11D^2+15DF-12F^2)\Gamma_2(m_K,w)+12(D+F)^2\Big(\Gamma_2(m_\pi,-w)\nonumber\\&-\Gamma_2(m_\pi,w)\Big)\Big]w+D(5D+3F)\Gamma_2(m_K,-w)(\delta+w)\nonumber\\&+2\Gamma_2(m_K,-\delta)\Big(-2\delta(4D^2-3DF+3F^2)+3(D-2F)(D-F)w\Big)\Bigg\}.
\end{align}
Fig. (17):
\begin{align}
\label{A33}
V_{(\pi N)}^{(3/2)}=&\frac{\mathcal{C}^2}{72f^4}(w^2-m_\pi^2+\frac{t}{2})\Bigg\{\frac{1}{\delta-w}\Bigg(-6\Big((D-3F)^2G_2(m_\eta,0)+8F^2G_2(m_K,0)\nonumber\\&+5D^2G_2(m_\pi,0)+10DFG_2(m_\pi,0)+5F^2G_2(m_\pi,0)-D^2\Gamma_2(m_\eta,-\delta+w)\nonumber\\&+6DF\Gamma_2(m_\eta,-\delta+w)-9F^2\Gamma_2(m_\eta,-\delta+w)-8F^2\Gamma_2(m_K,-\delta+w)\Big)\nonumber\\&+30(D+F)^2\Gamma_2(m_\pi,-\delta+w)\Bigg)+\frac{1}{\delta+w}\Big[-2(D-3F)^2G_2(m_\eta,0)\nonumber\\&-6(D+F)(D+3F)G_2(m_K,0)+7(D+F)G_2(m_\pi,0)\nonumber\\&+2(D-3F)^2\Gamma_2(m_\eta,-\delta-w)+6(D+F)(D+3F)\Gamma_2(m_K,-\delta-w)\nonumber\\&+42(D+F)^2\Gamma_2(m_\pi,-\delta-w)\Big]\Bigg\},
\end{align}
\begin{align}
\label{A34}
W_{(\pi N)}^{(3/2)}=&\frac{\mathcal{C}^2}{432f^4}\Bigg\{\frac{6}{\delta-w}\Big[-(D-3F)^2G_2(m_\eta,0)-5(D+F)^2G_2(m_\pi,0)\nonumber\\&+D^2\Gamma_2(m_\eta,-\delta+w)-6DF\Gamma_2(m_\eta,-\delta+w)+9F^2\Gamma_2(m_\eta,-\delta+w)\nonumber\\&+8F^2\Gamma_2(m_K,-\delta+w)+5(D+F)^2\Gamma_2(m_\pi,-\delta+w)\Big]\nonumber\\&+\frac{2}{\delta+w}\Big[(D-3F)^2G_2(m_\eta,0)+21(D+F)^2G_2(m_\pi,0)-D^2\Gamma_2(m_\eta,-\delta-w)\nonumber\\&+6DF\Gamma_2(m_\eta,-\delta-w)-9F^2\Gamma_2(m_\eta,-\delta-w)-3D^2\Gamma_2(m_K,-\delta-w)\nonumber\\&-12DF\Gamma_2(m_K,-\delta-w)-9F^2\Gamma_2(m_K,-\delta-w)-21(D\nonumber\\&+F)^2\Gamma_2(m_\pi,-\delta-w)\Big]+6G_2(m_K,0)\Bigg(\frac{8F^2}{-\delta+w}+\frac{(D+F)(D+3F)}{\delta+w}\Bigg)\Bigg\},
\end{align}
\begin{align}
\label{A35}
V_{(\pi N)}^{(1/2)}=&-\frac{\mathcal{C}^2(-2m_\pi^2+t+2w^2)}{144f^4(\delta^2-w^2)}\Bigg\{\delta\Bigg[72(D+F)^2G_2(m_\pi,0)-2(D-3F)(4D\nonumber\\&+15F)\Gamma_2(m_\eta,-\delta+w)+3\Big(6F(-D+3F)\Gamma_2(m_\eta,-\delta+w)+(D-3F)(D\nonumber\\&+7F)\Gamma_2(m_K,-\delta-w)-9D^2\Gamma_2(m_K,-\delta+w)-6DF\Gamma_2(m_K,-\delta+w)\nonumber\\&-F^2\Gamma_2(m_K,-\delta+w)-8D^2\Gamma_2(m_\pi,-\delta-w)-16DF\Gamma_2(m_\pi,-\delta-w)\nonumber\\
&-8F^2\Gamma_2(m_\pi,-\delta-w)-16(D+F)^2\Gamma_2(m_\pi,-\delta+w)\Big)\Bigg]+\Bigg[24(D+F)^2\nonumber\\&G_2(m_\pi,0)+2(D-3F)(4D+15F)\Gamma_2(m_\eta,-\delta-w)+3(D-3F)\nonumber\\&\Big(-6F\Gamma_2(m_\eta,-\delta+w)-(D+7F)\Gamma_2(m_K,-\delta-w)\Big)-3(3D+F)^2\nonumber\\&\Gamma_2(m_K,-\delta+w)+24(D+F)^2\Gamma_2(m_\pi,-\delta-w)-48(D+F)^2\nonumber\\&\Gamma_2(m_\pi,-\delta+w)\Bigg]w+6G_2(m_K,0)\Big[\delta(4D^2+DF+11F^2)+5(D-F)(D\nonumber\\&+2F)w\Big]+4(D-3F)G_2(m_\eta,0)\Big[2\delta(D+6F)-(2D+3F)w\Big]\Bigg\},
\end{align}
\begin{align}
\label{A36}
W_{(\pi N)}^{(1/2)}=&\frac{\mathcal{C}^2}{432f^4(\delta^2-w^2)}\Bigg\{\delta\Bigg[-24(D+F)^2G_2(m_\pi,0)-2(D-3F)(4D+15F)\nonumber\\&\Gamma_2(m_\eta,-\delta-w)+3\Big(6(D-3F)F\Gamma_2(m_\eta,-\delta+w)+(D-3F)(D\nonumber\\&+7F)\Gamma_2(m_K,-\delta-w)+9D^2\Gamma_2(m_K,-\delta+w)+6DF\Gamma_2(m_K,-\delta+w)\nonumber\\&+F^2\Gamma_2(m_K,-\delta+w)-8D^2\Gamma_2(m_\pi,-\delta-w)-16DF\Gamma_2(m_\pi,-\delta-w)\nonumber\\
&-8F^2\Gamma_2(m_\pi,-\delta-w)+16(D+F)^2\Gamma_2(m_\pi,-\delta+w)\Big)\Bigg]+\Bigg[-72(D+F)^2\nonumber\\&G_2(m_\pi,0)+2(D-3F)(4D+15F)\Gamma_2(m_\eta,-\delta-w)+3\Big(6(D-3F)F\nonumber\\&\Gamma_2(m_\eta,-\delta+w)-(D-3F)(D+7F)\Gamma_2(m_K,-\delta-w)+9D^2\Gamma_2(m_K,-\delta+w)\nonumber\\&+6DF\Gamma_2(m_K,-\delta+w)+F^2\Gamma_2(m_
K,-\delta+w)+8D^2\Gamma_2(m_\pi,-\delta-w)\nonumber\\&+16DF\Gamma_2(m_\pi,-\delta-w)+8F^2\Gamma_2(m_\pi,-\delta-w)\nonumber\\&+16(D+F)^2\Gamma_2(m_\pi,-\delta+w)\Big)\Bigg]w+4(D-3F)G_2(m_\eta,0)\Big[\delta(2D+3F)-2(D\nonumber\\&+6F)w\Big]+6G_2(m_K,0)\Big[-5\delta(D-F)(D+2F)-(4D^2+DF+11F^2)w\Big]\Bigg\}.
\end{align}

Figs. (18)+(19):
\begin{align}
\label{A37}
V_{(\pi N)}^{(3/2)}=0,
\end{align}
\begin{align}
\label{A38}
W_{(\pi N)}^{(3/2)}=&\frac{5\mathcal{C}^2\mathcal{H}}{81wf^4}\Big[2(-D+F)\Gamma_2(m_K,-\delta)+2D\Gamma_2(m_K,-\delta-w)\nonumber\\&-2F\Gamma_2(m_K,-\delta+w)+5(D+F)\Big(\Gamma_2(m_\pi,-\delta-w)-\Gamma_2(m_\pi,-\delta+w)\Big)\Big],
\end{align}
\begin{align}
\label{A39}
V_{(\pi N)}^{(1/2)}=0,
\end{align}
\begin{align}
\label{A40}
W_{(\pi N)}^{(1/2)}=&\frac{5\mathcal{C}^2\mathcal{H}}{81wf^4}\Big[2(-D+F)\Gamma_2(m_K,-\delta)-(D+3F)\Gamma_2(m_K,-\delta-w)\nonumber\\&+3D\Gamma_2(m_K,-\delta+w)+F\Gamma_2(m_K,-\delta+w)-10D\Gamma_2(m_\pi,-\delta-w)\nonumber\\&-10F\Gamma_2(m_\pi,-\delta-w)+10(D+F)\Gamma_2(m_\pi,-\delta+w)\Big].
\end{align}
Fig. (20):
\begin{align}
\label{A41}
V_{(\pi N)}^{(3/2)}=&\frac{\mathcal{C}^4}{216f^4}(w^2-m_\pi^2+\frac{t}{2})\Bigg\{\frac{1}{(\delta-w)^2}\Big[9J_2(m_K,-\delta)-9J_2(m_K,-w)\nonumber\\&+44\Big(J_2(m_\pi,-\delta)-J_2(m_\pi,-w)\Big)\Big]+\frac{1}{\delta-w}\Big[9G_2(m_K,-\delta)\nonumber\\&+44G_2(m_\pi,-\delta)\Big]+\frac{1}{(\delta+w)^2}\Big[J_2(m_K,-\delta)-J_2(m_K,w)+4J_2(m_\pi,-\delta)\nonumber\\&-4J_2(m_\pi,w)\Big]+\frac{1}{\delta+w}\Big[G_2(m_K,-\delta)+4G_2(m_\pi,-\delta)\Big]\Bigg\},
\end{align}
\begin{align}
\label{A42}
W_{(\pi N)}^{(3/2)}=&\frac{5\mathcal{C}^4}{1296f^4}\Bigg\{\frac{1}{(\delta-w)^2}\Big[-9J_2(m_K,-\delta)+9J_2(m_K,-w)+44\Big(-J_2(m_\pi,-\delta)\nonumber\\&+J_2(m_\pi,-w)\Big)\Big]+\frac{1}{-\delta+w}\Big[9G_2(m_K,-\delta)+44G_2(m_\pi,-\delta)\Big]\nonumber\\&+\frac{1}{(\delta+w)^2}\Big[J_2(m_K,-\delta)-J_2(m_K,w)+4J_2(m_\pi,-\delta)-4J_2(m_\pi,w)\Big]\nonumber\\&+\frac{1}{\delta+w}\Big[G_2(m_K,-\delta)+4G_2(m_\pi,-\delta)\Big]\Bigg\},
\end{align}
\begin{align}
\label{A43}
V_{(\pi N)}^{(1/2)}=&\frac{\mathcal{C}^4(2w^2-2m_\pi^2+t)}{432f^4(\delta^2-w^2)^2}\Bigg\{\delta^2\Bigg[10J_2(m_K,-\delta)+3J_2(m_K,-w)-13J_2(m_K,w)\nonumber\\&+16\Big(3J_2(m_\pi,-\delta)+J_2(m_\pi,-w)-4J_2(m_\pi,w)\Big)\Bigg]+2\delta\Big(-16J_2(m_K,-\delta)\nonumber\\&+3J_2(m_K,-w)+13J_2(m_K,w)+16(-5J_2(m_\pi,-\delta)+J_2(m_\pi,-w)\nonumber\\&+4J_2(m_\pi,w))\Big)w+\Bigg[10J_2(m_K,-\delta)+3J_2(m_K,-w)-13J_2(m_K,w)\nonumber\\&+16\Big(3J_2(m_\pi,-\delta)+J_2(m_\pi,-w)-4J_2(m_\pi,w)\Big)\Bigg]w^2\nonumber\\&+2G_2(m_K,-\delta)(5\delta-8w)(\delta^2-w^2)+16G_2(m_\pi,-\delta)(3\delta-5w)(\delta^2-w^2)\Bigg\},
\end{align}
\begin{align}
\label{A44}
W_{(\pi N)}^{(1/2)}=&\frac{5\mathcal{C}^4}{1296f^4}\Bigg\{\frac{1}{(\delta-w)^2}\Big[3J_2(m_K,-\delta)-3J_2(m_K,-w)+16\Big(J_2(m_\pi,-\delta)-J_2(m_\pi,-w)\Big)\Big]\nonumber\\&+\frac{1}{\delta-w}\Big[3G_2(m_K,-\delta)+16G_2(m_\pi,-\delta)\Big]+\frac{1}{(\delta+w)^2}\Big[13J_2(m_K,-\delta)-13J_2(m_K,w)\nonumber\\&+64\Big(J_2(m_\pi,-\delta)-J_2(m_\pi,w)\Big)\Big]+\frac{1}{\delta+w}\Big[13G_2(m_K,-\delta)+64G_2(m_\pi,-\delta)\Big]\Bigg\}.
\end{align}

Fig. (21):
\begin{align}
\label{A45}
V_{(\pi N)}^{(3/2)}=&-\frac{5\mathcal{C}^2\mathcal{H}^2}{972f^4w^2}(-2m_\pi^2+t+2w^2)\Big[-4J_2(m_K,-\delta)+3J_2(m_K,-\delta-w)\nonumber\\&+J_2(m_K,-\delta+w)-30J_2(m_\pi,-\delta)+20J_2(m_\pi,-\delta-w)\nonumber\\&+10J_2(m_\pi,-\delta+w)+2\Big(G_2(m_K,-\delta)+5G_2(m_\pi,-\delta)\Big)w\Big],
\end{align}
\begin{align}
\label{A46}
W_{(\pi N)}^{(3/2)}=&-\frac{5\mathcal{C}^2\mathcal{H}^2}{1458f^4w^2}\Big[2J_2(m_K,-\delta)-3J_2(m_K,-\delta-w)+J_2(m_K,-\delta+w)\nonumber\\&+10\Big(J_2(m_\pi,-\delta)-2J_2(m_\pi,-\delta-w)+J_2(m_\pi,-\delta+w)\Big)\nonumber\\&-2\Big(2G_2(m_K,-\delta)+15G_2(m_\pi,-\delta)\Big)w\Big],
\end{align}
\begin{align}
\label{A47}
V_{(\pi N)}^{(1/2)}=&\frac{5\mathcal{C}^2\mathcal{H}^2}{972f^4w^2}(-2m_\pi^2+t+2w^2)\Big[4J_2(m_K,-\delta)-4J_2(m_K,-\delta+w)\nonumber\\&+30J_2(m_\pi,-\delta)-5\Big(J_2(m_\pi,-\delta-w)+5J_2(m_\pi,-\delta+w)\Big)\nonumber\\&+4\Big(G_2(m_K,-\delta)+5G_2(m_\pi,-\delta)\Big)w\Big],
\end{align}
\begin{align}
\label{A48}
W_{(\pi N)}^{(1/2)}=&\frac{5\mathcal{C}^2\mathcal{H}^2}{1458f^4w^2}\Big[4J_2(m_K,-\delta)-4J_2(m_K,-\delta+w)\nonumber\\&+5\Big(4J_2(m_\pi,-\delta)+J_2(m_\pi,-\delta-w)-5J_2(m_\pi,-\delta+w)\Big)\nonumber\\&+2\Big(2G_2(m_K,-\delta)+15G_2(m_\pi,-\delta)\Big)w\Big].
\end{align}
Fig. (22):
\begin{align}
\label{A49}
V_{(\pi N)}^{(3/2)}=-\frac{\mathcal{C}^2(D+F)^2}{8f^4w^2}(-2m_\pi^2+t+2w^2)\Big[J_2(m_K,-\delta-w)+4J_2(m_\pi,-\delta-w)\Big],
\end{align}
\begin{align}
\label{A50}
W_{(\pi N)}^{(3/2)}=\frac{\mathcal{C}^2(D+F)^2}{4f^4w^2}\Big[J_2(m_K,-\delta-w)+4J_2(m_\pi,-\delta-w)\Big],
\end{align}
\begin{align}
\label{A51}
V_{(\pi N)}^{(1/2)}=&\frac{\mathcal{C}^2(D+F)^2}{16f^4w^2}(-2m_\pi^2+t+2w^2)\Big[J_2(m_K,-\delta-w)-3J_2(m_K,-\delta+w)\nonumber\\&+4\Big(J_2(m_\pi,-\delta-w)-3J_2(m_\pi,-\delta+w)\Big)\Big],
\end{align}
\begin{align}
\label{A52}
W_{(\pi N)}^{(1/2)}=&-\frac{\mathcal{C}^2(D+F)^2}{8f^4w^2}\Big[J_2(m_K,-\delta-w)-3J_2(m_K,-\delta+w)\nonumber\\&+4\Big(J_2(m_\pi,-\delta-w)+3J_2(m_\pi,-\delta+w)\Big)\Big].
\end{align}
Fig. (23):
\begin{align}
\label{A53}
V_{(\pi N)}^{(3/2)}=-\frac{\mathcal{C}^4}{6f^4}(-m_\pi^2+\frac{t}{2}+w^2)\Bigg[\frac{J_2(m_K,w)+J_2(m_\pi,w)}{(\delta-w)^2}+\frac{J_2(m_K,-w)+J_2(m_\pi,-w)}{3(\delta+w)^2}\Bigg],
\end{align}
\begin{align}
\label{A54}
W_{(\pi N)}^{(3/2)}=\frac{\mathcal{C}^4}{12f^4}\Bigg[\frac{J_2(m_K,w)+J_2(m_\pi,w)}{(\delta-w)^2}-\frac{J_2(m_K,-w)+J_2(m_\pi,-w)}{3(\delta+w)^2}\Bigg],
\end{align}
\begin{align}
\label{A55}
V_{(\pi N)}^{(1/2)}=-\frac{\mathcal{C}^4}{9f^4(\delta+w)^2}(-2m_\pi^2+t+2w^2)\Big[J_2(m_K,-w)+J_2(m_\pi,-w)\Big],
\end{align}
\begin{align}
\label{A56}
W_{(\pi N)}^{(1/2)}=-\frac{\mathcal{C}^4}{9f^4(\delta+w)^2}\Big[J_2(m_K,-w)+J_2(m_\pi,-w)\Big].
\end{align}
Fig. (24):
\begin{align}
\label{A57}
V_{(\pi N)}^{(3/2)}=&\frac{5\mathcal{C}^2\mathcal{H}^2}{324f^4}(-m_\pi^2+\frac{t}{2}+w^2)\Bigg\{-\frac{3}{(\delta-w)^2}\Big[J_2(m_\eta,-\delta+w)+2J_2(m_K,-\delta+w)\nonumber\\&+5J_2(m_\pi,-\delta+w)\Big]-\frac{1}{(\delta+w)^2}\Big[J_2(m_\eta,-\delta-w)+2J_2(m_K,-\delta-w)\nonumber\\&+5J_2(m_\pi,-\delta-w)\Big]\Bigg\},
\end{align}
\begin{align}
\label{A58}
W_{(\pi N)}^{(3/2)}=&\frac{5\mathcal{C}^2\mathcal{H}^2}{648f^4}\Bigg\{\frac{3}{(\delta-w)^2}\Big[J_2(m_\eta,-\delta+w)+2J_2(m_K,-\delta+w)+5J_2(m_\pi,-\delta+w)\Big]\nonumber\\&-\frac{1}{(\delta+w)^2}\Big[J_2(m_\eta,-\delta-w)+2J_2(m_K,-\delta-w)+5J_2(m_\pi,-\delta-w)\Big]\Bigg\},
\end{align}
\begin{align}
\label{A59}
V_{(\pi N)}^{(1/2)}=&-\frac{5\mathcal{C}^2\mathcal{H}^2}{162f^4(\delta+w)^2}(-2m_\pi^2+t+2w^2)\Big[J_2(m_\eta,-\delta-w)+2J_2(m_K,-\delta-w)\nonumber\\&+5J_2(m_\pi,-\delta-w)\Big],
\end{align}
\begin{align}
\label{A60}
W_{(\pi N)}^{(1/2)}=-\frac{5\mathcal{C}^2\mathcal{H}^2}{162f^4(\delta+w)^2}\Big[J_2(m_\eta,-\delta-w)+2J_2(m_K,-\delta-w)+5J_2(m_\pi,-\delta-w)\Big].
\end{align}
Figs. (25)+(26):
\begin{align}
\label{A61}
V_{(\pi N)}^{(3/2)}=-\frac{\mathcal{C}^2(2\delta+w)}{216f^4\pi^2(\delta^2-w^2)}(-2m_\pi^2+t+2w^2)\Bigg(m_K^2\text{ln}\frac{m_K}{\lambda}+2m_\pi^2\text{ln}\frac{m_\pi}{\lambda}\Bigg),
\end{align}
\begin{align}
\label{A62}
W_{(\pi N)}^{(3/2)}=\frac{\mathcal{C}^2(\delta+2w)}{216f^4\pi^2(\delta^2-w^2)}\Bigg(m_K^2\text{ln}\frac{m_K}{\lambda}+2m_\pi^2\text{ln}\frac{m_\pi}{\lambda}\Bigg),
\end{align}
\begin{align}
\label{A63}
V_{(\pi N)}^{(1/2)}=-\frac{\mathcal{C}^2}{108f^4\pi^2(\delta+w)}(-2m_\pi^2+t+2w^2)\Bigg(m_K^2\text{ln}\frac{m_K}{\lambda}+2m_\pi^2\text{ln}\frac{m_\pi}{\lambda}\Bigg),
\end{align}
\begin{align}
\label{A64}
W_{(\pi N)}^{(1/2)}=-\frac{\mathcal{C}^2}{108f^4\pi^2(\delta+w)}\Bigg(m_K^2\text{ln}\frac{m_K}{\lambda}+2m_\pi^2\text{ln}\frac{m_\pi}{\lambda}\Bigg).
\end{align}
Figs. (27)+(28):
\begin{align}
\label{A65}
V_{(\pi N)}^{(3/2)}=\frac{\mathcal{C}^2}{36f^4}\Bigg[3J_2(m_K,-\delta)+32J_2(m_\pi,-\delta)\Bigg],
\end{align}
\begin{align}
\label{A66}
W_{(\pi N)}^{(3/2)}=0,
\end{align}
\begin{align}
\label{A67}
V_{(\pi N)}^{(1/2)}=\frac{\mathcal{C}^2}{36f^4}\Bigg[3J_2(m_K,-\delta)+32J_2(m_\pi,-\delta)\Bigg],
\end{align}
\begin{align}
\label{A68}
W_{(\pi N)}^{(1/2)}=0.
\end{align}
Fig. (29):
\begin{align}
\label{A69}
V_{(\pi N)}^{(3/2)}=&\frac{\mathcal{C}^2}{108f^4}\Bigg\{-84J_2(m_\pi,-\delta)-(12m_\pi^2-7t)\Big[2J_0^a(m_\pi)+2{\delta}J_0^F(m_\pi)\nonumber\\&-J_0^T(m_\pi)(2\delta^2-2m_\pi^2+t)\Big]-3\Big[6{\delta}J_0^a(m_\pi)-8J_0^C(m_\pi)\nonumber\\&+3{\delta}J_0^T(m_\pi)\Big(4(\delta-m_\pi)(\delta+m_\pi)-t\Big)\nonumber\\&+4J_0^F(m_\pi)(-3{\delta}^2+2m_\pi^2+t)\Big]w\Bigg\},
\end{align}
\begin{align}
\label{A70}
W_{(\pi N)}^{(3/2)}=\frac{\mathcal{C}^2}{24f^4}\Big[2J_0^a(m_\pi)+4{\delta}J_0^F(m_\pi)-J_0^T(m_\pi)(4\delta^2-4m_\pi^2+t)\Big],
\end{align}
\begin{align}
\label{A71}
V_{(\pi N)}^{(1/2)}=&\frac{\mathcal{C}^2}{108f^4}\Bigg\{-84J_2(m_\pi,-\delta)-(12m_\pi^2-7t)\Big[2J_0^a(m_\pi)+2{\delta}J_0^F(m_\pi)\nonumber\\&-J_0^T(m_\pi)(2\delta^2-2m_\pi^2+t)\Big]-3\Big[6{\delta}J_0^a(m_\pi)-8J_0^C(m_\pi)\nonumber\\&+3{\delta}J_0^T(m_\pi)\Big(4(\delta-m_\pi)(\delta+m_\pi)-t\Big)\nonumber\\&+4J_0^F(m_\pi)(-3{\delta}^2+2m_\pi^2+t)\Big]w\Bigg\},
\end{align}
\begin{align}
\label{A72}
W_{(\pi N)}^{(1/2)}=\frac{\mathcal{C}^2}{24f^4}\Big[2J_0^a(m_\pi)+4{\delta}J_0^F(m_\pi)-J_0^T(m_\pi)(4\delta^2-4m_\pi^2+t)\Big].
\end{align}
Fig. (30):
\begin{align}
\label{A73}
V_{(\pi N)}^{(3/2)}=-\frac{\mathcal{C}^2w}{3f^4}\Bigg[G_2(m_K,-\delta)+5G_2(m_\pi,-\delta)\Bigg],
\end{align}
\begin{align}
\label{A74}
W_{(\pi N)}^{(3/2)}=0,
\end{align}
\begin{align}
\label{A75}
V_{(\pi N)}^{(1/2)}=\frac{2\mathcal{C}^2w}{3f^4}\Bigg[G_2(m_K,-\delta)+5G_2(m_\pi,-\delta)\Bigg],
\end{align}
\begin{align}
\label{A76}
W_{(\pi N)}^{(1/2)}=0.
\end{align}
For giving the expressions as much detail as possible, we use several functions in the obove amplitudes. They are given by 
\begin{align}
\label{A77}
\Delta(m)=\frac{m^2}{8\pi^2}\text{ln}\frac{m}{\lambda},
\end{align}
\begin{equation}
\label{A78}
J_0(m,w)=\frac{w}{8\pi^2}(1-2\text{ln}\frac{m}{\lambda})
+\left\{
\begin{array}{ll}
-\frac{\sqrt{m^2-w^2}}{4\pi^2}\arccos{\frac{-w}{m}}&  -m < w < m \\
\\
\frac{\sqrt{w^2-m^2}}{4\pi^2}\text{ln}\frac{\sqrt{w^2-m^2}-w}{m}  &  w < -m ,\\
\\
\frac{\sqrt{w^2-m^2}}{4\pi^2}(i\pi-\text{ln}\frac{\sqrt{w^2-m^2}+w}{m}) &  w > m \\
\end{array} \right.
\end{equation}
% \begin{align}
%\label{A78}
%J_1(m,w)=wJ_0(m,w)+\frac{m^2}{8\pi^2}\text{ln}\frac{m}{\lambda},
%\end{align}
\begin{align}
\label{A79}
J_2(m,w)=\frac{1}{3}\Bigg\{(m^2-w^2)J_0(m,w)-\frac{wm^2}{8\pi^2}\text{ln}\frac{m}{\lambda}+\frac{1}{8\pi^2}\Big[wm^2-\frac{2}{3}w^3\Big]\Bigg\},
\end{align}
%\begin{align}
%\label{A80}
%J_3(m,w)=wJ_1(m,w)-J_2(m,w),
%\end{align}
\begin{align}
\label{A80}
G_2(m,w)=\frac{ \partial  }{ \partial w }
J_2(m,w),
\end{align}
\begin{align}
\label{A81}
\Gamma_2(m,w)=\frac{1}{w}
\Big[J_2(m,w)-J_2(m,0)\Big],
\end{align}
\begin{align}
\label{A82}
J^C_0(m)=\Delta(m),
\end{align}
\begin{equation}
\label{A83}
J^a_0(m)=\frac{1}{8\pi^2}\Big[\pi\sqrt{m^2-\delta^2}+
\Big(\delta\text{ln}\frac{m^2}{\lambda^2}-\delta\Big)\Big]+\left\{
\begin{array}{ll}
\frac{\sqrt{m^2-\delta^2}}{4\pi^2}\arctan{\frac{\delta}{\sqrt{m^2-\delta^2}}}& m^2 > \delta^2 \\
\\
\frac{\sqrt{\delta^2-m^2}}{8\pi^2}\Big(i\pi-\text{ln}\frac{\delta-\sqrt{\delta^2-m^2}}{\delta+\sqrt{\delta^2-m^2}}\Big)& m^2 < \delta^2 
\end{array} \right.,
\end{equation}
\begin{align}
\label{A84}
J^F_0(m)=-\frac{1}{16\pi^2}\Bigg(1-\text{ln}\frac{m^2}{\lambda^2}-r\text{ln}\Big\lvert\frac{1+r}{1-r}\Big\rvert\Bigg),\Bigg(r=\sqrt{\Big\lvert1-\frac{4m^2}{t}\Big\rvert}\Bigg),
\end{align}
\begin{align}
\label{A85}
J^T_0(m)=\frac{1}{16\pi^2\sqrt{-t}}\Big[F_1(y_1)+F_2(x_1)-F_3(z_1)+F_1(y_2)+F_2(x_2)-F_3(z_2)\Big]
\end{align}
with 
\begin{align}
\label{A86}
F_1(x)=&\text{Li}_2(\frac{z_0+2t}{z_0-2x\sqrt{-t}})-\text{Li}_2(\frac{z_0}{z_0-2x\sqrt{-t}}),\nonumber\\
F_2(x)=&-\text{Li}_2(\frac{z_0+2t}{z_0-2x\sqrt{-t}})-\frac{1}{2}\text{ln}^2(\frac{z_0-2x\sqrt{-t}}{\lambda^2}),\nonumber\\
F_3(x)=&F_1(x)+F_2(x),\nonumber\\
x_{1,2}=&\sqrt{-t}+\delta\pm\sqrt{\delta^2-m^2},\quad
y_{1,2}=-\frac{1}{2\sqrt{-t}}\Big[t\pm\sqrt{(t-4m^2)t}\Big],\nonumber\\
z_{1,2}=&\delta\pm\sqrt{\delta^2-m^2},\quad
z_0=\sqrt{-t}(2\delta+\sqrt{-t}).
\end{align}
The dilogarithm or Spence function (polylogarithm function) is defined by
\begin{align}
\label{A87}
\text{Li}_2(x)=-\int_{0}^{1}\frac{\text{ln}(1-xt)}{t},
\end{align}

External legs (wave function) without decuplet renormalization:
\begin{align}
\label{A88}
V_{(\pi N)}^{(3/2)}=\frac{2\mathcal{C}^2}{9f^4(\delta^2-w^2)}(2\delta+w)(-2m_\pi^2+t+2w^2)(\delta_{Z_\pi}+\delta_{Z_N}),
\end{align}
\begin{align}
\label{A89}
W_{(\pi N)}^{(3/2)}=-\frac{2\mathcal{C}^2}{9f^4(\delta^2-w^2)}(\delta_{Z_\pi}+\delta_{Z_N})(2\delta+w),
\end{align}
\begin{align}
\label{A90}
V_{(\pi N)}^{(1/2)}=\frac{4\mathcal{C}^2}{9f^4(\delta+w)}(\delta_{Z_\pi}+\delta_{Z_N})(-2m_\pi^2+t+2w^2),
\end{align}
\begin{align}
\label{A91}
W_{(\pi N)}^{(1/2)}=-\frac{4\mathcal{C}^2}{9f^4(\delta+w)}(\delta_{Z_\pi}+\delta_{Z_N}),
\end{align}
where
\label{A92}
\begin{align}
\delta_{Z_\pi}=\frac{1}{24\pi^2}\Big[2m_\pi^2\text{ln}\frac{m_\pi}{\lambda}+m_K^2\text{ln}\frac{m_K}{\lambda}\Big],
\end{align}
\label{A93}
\begin{align}
\delta_{Z_N}=&-\frac{1}{96\pi^2}\Bigg\{9(D+F)^2 m_\pi^2\Big[1+3\text{ln}\frac{m_\pi}{\lambda}\Big]+(10D^2+18F^2\nonumber\\&-12DF)m_K^2\Big[1+3\text{ln}\frac{m_K}{\lambda}\Big]+(D-3F)^2m_\eta^2\Big[1+3\text{ln}\frac{m_\eta}{\lambda}\Big]\Bigg\}.
\end{align}
External legs (wave function) with decuplet renormalization:
\begin{align}
\label{A94}
V_{(\pi N)}^{(3/2)}=\frac{1}{2wf^4}\Big[(D+F)^2(2w^2-2m_\pi^2+t)-w^2\Big]\delta_{Z_{NT}},
\end{align}
\begin{align}
\label{A95}
W_{(\pi N)}^{(3/2)}=-\frac{(D+F)^2}{wf^4}\delta_{Z_{NT}},
\end{align}
\begin{align}
\label{A96}
V_{(\pi N)}^{(1/2)}=\frac{1}{wf^4}\Big[-(D+F)^2(2w^2-2m_\pi^2+t)+2w^2\Big]\delta_{Z_{NT}},
\end{align}
\begin{align}
\label{A97}
W_{(\pi N)}^{(1/2)}=-\frac{(D+F)^2}{wf^4}\delta_{Z_{NT}},
\end{align}
where
\label{A98}
\begin{align}
\delta_{Z_{NT}}=\frac{7\mathcal{C}^2}{96\pi^2}Q(m_\pi)-\frac{\mathcal{C}^2}{288\pi^2}Q(m_\eta)
\end{align}
with 
\label{A99}
\begin{equation}
W(m)=\delta\text{ln}\frac{\lvert{m}\rvert}{\lambda}+\left\{
\begin{array}{ll}
-\sqrt{m^2-\delta^2}\arccos{\frac{\delta}{m}}& m^2 > \delta^2 \\
\\
+\sqrt{\delta^2-m^2}\text{ln}\frac{\delta+\sqrt{\delta^2-m^2}}{\lvert{m}\rvert}& m^2 < \delta^2 
\end{array} \right.,
\end{equation}
\label{A100}
\begin{align}
Q(m)=m^2\Big(1+2\text{ln}\frac{\lvert{m}\rvert}{\lambda}\Big)-4\delta{W(m)}.
\end{align}

\newpage
\bibliographystyle{utphys}
\bibliography{pion_nucleon_scattering_with_decuplet}% Produces the bibliography via BibTeX.
\end{document}